\documentclass[11pt]{article}
\pdfoutput=1
\usepackage{soul}
\usepackage{jheppub}
\usepackage{amsfonts}
\usepackage{amsmath}
\usepackage{amsthm}
\allowdisplaybreaks[4]         
\usepackage{amssymb}   
\usepackage{euscript}   
\usepackage{cleveref}    
\usepackage[dvipsnames]{xcolor}          
\usepackage{tensor}     
\usepackage{graphicx}
\usepackage{caption}
\usepackage{array}
\usepackage{booktabs}
\usepackage{subcaption}
\usepackage{multirow}
\usepackage[skip=0.5ex, belowskip=1ex,
                labelformat=brace,
                singlelinecheck=off]{subcaption}   
\usepackage{float}
\usepackage{enumitem}
\usepackage{hyperref}
\hypersetup{
	colorlinks =NavyBlue,
	linkcolor =NavyBlue,
	citecolor=MidnightBlue,
	filecolor=NavyBlue,
	urlcolor=NavyBlue,
}

\newcommand{\bea}{\begin{eqnarray}}
\newcommand{\eea}{\end{eqnarray}}
\newcommand{\ba}{\begin{eqnarray}}
\usepackage{braket}
\newcommand{\ea}{\end{eqnarray}}

\newcommand{\beq}{\begin{equation}}
\newcommand{\eeq}{\end{equation} }
\newcommand{\beqa}{\begin{eqnarray}}

\newcommand{\eeqa}{\end{eqnarray}}
\newcommand{\beqar}{\begin{eqnarray*}}
\newcommand{\eeqar}{\end{eqnarray*}}

\newcommand{\be}{\begin{equation}}
\newcommand{\ee}{\end{equation}}

\newtheorem*{defi*}{Definition}

\usepackage{tikz}
\usetikzlibrary{mindmap,trees,shadows,backgrounds}
\usepackage{tikz-3dplot}
\usetikzlibrary{positioning,calc}

\tikzset{
    invisible/.style={opacity=0},
    visible on/.style={alt={#1{}{invisible}}},
    alt/.code args={<#1>#2#3}{%
      \alt<#1>{\pgfkeysalso{#2}}{\pgfkeysalso{#3}} 
    },
  }

\definecolor{shadecolor}{rgb}{.25,.25,.25}

\preprint{\texttt{WI-07-2026, IFT-UAM/CSIC-26-20}}

\title{ \boldmath Higher-dimensional BKL dynamics in AdS black holes}

\author[a]{Elena C\'aceres,}
\author[b]{\'Angel J. Murcia,}
\author[c]{Ayan K. Patra,}
\author[d]{Juan F. Pedraza}

\affiliation[a]{Theory Group, Department of Physics, University of Texas, Austin, TX 78712, USA
 \vspace{0.1cm}}
\affiliation[b]{Departament de F\'isica Qu\`antica i Astrof\'isica, Institut de Ci\`encies del Cosmos\\ Universitat de Barcelona, Mart\'i i Franqu\'es 1, E-08028 Barcelona, Spain \vspace{0.1cm}}
\affiliation[c]{Centre for Particle Theory, Department of Mathematical Sciences, Durham University,\\ Durham DH1 3LE, UK\vspace{0.1cm}}
\affiliation[d]{Instituto de Física Teórica UAM/CSIC, Calle Nicol\'as Cabrera 13-15, Madrid, E-28049, Spain}

\emailAdd{elenac@utexas.edu}
\emailAdd{angelmurcia@icc.ub.edu}
\emailAdd{ayan-kumar.patra@durham.ac.uk}
\emailAdd{j.pedraza@csic.es}

\abstract{Chaotic BKL dynamics provides a canonical description of the approach to spacelike singularities as a sequence of Kasner epochs grouped into eras. While this paradigm is well established for cosmological singularities, explicit realizations inside black holes have been scarce, despite renewed interest from holography. Here, we construct a broad class of asymptotically AdS black holes in $D\ge 4$ whose interiors exhibit bona fide BKL dynamics as the singularity is approached. In the near-singularity regime, the evolution reduces to billiard-like motion in a compact domain that forms a regular $(D-2)$-simplex. We derive closed-form bouncing rules for the Kasner exponents in arbitrary dimension and prove the ensuing chaotic dynamics. A key novelty for $D\ge 5$ is a richer internal organization of eras: inequivalent transitions between epochs lead to distinct \emph{Kasner seasons}, yielding new patterns of epoch/era structure for both electric and gravitational walls. Finally, we investigate a holographic diagnostic, the thermal $a$-function, whose monotonic flow captures individual epochs and eras and can display near-walking behavior in suitable Kasner regimes. 

}

\begin{document} 
\maketitle
\flushbottom

\section{Introduction}

Ever since the celebrated works of Penrose and Hawking~\cite{Penrose:1964wq,Hawking:1970zqf,Hawking:1973uf}, the presence of singularities has been established as a generic feature of solutions to the Einstein field equations. This realization sparked a sustained program to understand the structure of spacetime in the vicinity of singularities. Among its most striking outcomes is the emergence of chaotic dynamics near spacelike singularities~\cite{Misner1969,Belinsky:1970ew,Lifshitz:1971fas}. It was shown that, in four spacetime dimensions, the approach to a spacelike singularity in general relativity (GR) is ultralocal: the evolution of neighboring spatial points decouples.\footnote{While ultralocality was originally argued to be a feature of near-singularity dynamics, it has only recently been proven in a covariant expansion of GR based on Carrollian limits~\cite{Oling:2024vmq}, providing a systematic explanation for the generic emergence of chaotic dynamics.} The dynamics then proceeds through a sequence of regimes in which the metric is well approximated by a Kasner solution~\cite{kasner1921geometrical} , the so-called \emph{Kasner epochs}. Moreover, chaotic behavior naturally arises once one groups consecutive epochs into \emph{Kasner eras}. This chaotic regime is known as BKL (Belinski--Khalatnikov--Lifshitz) or Mixmaster dynamics~\cite{Misner1969,Belinsky:1970ew,Lifshitz:1971fas}. Besides its central role in our understanding of singularities, BKL dynamics has recently attracted renewed attention in the AdS/CFT community because of its potential implications for the dual field theory. The goal of this work is to characterize BKL dynamics in higher dimensions and to provide explicit, asymptotically AdS, black hole solutions that realize it. In the process, we identify that Kasner epochs may feature different \emph{Kasner seasons}, a novel notion that arises non-trivially in dimensions $D\geq 5$.

The near-singularity behavior of solutions in higher-dimensional GR was originally explored in~\cite{Demaret:1985jnc,Demaret:1986ys,Elskens:1987gj,Elskens:1987rk}. Interestingly, in the absence of matter, the appearance of (generalized) BKL dynamics depends on the number of spacetime dimensions: purely gravitational systems exhibit oscillatory/chaotic behavior only for $D\leq 10$. Nevertheless, adding matter---for instance, $p$-form field strengths---generically restores chaos near spacelike singularities~\cite{Damour:2000wm,Damour:2000th}, allowing for chaotic BKL dynamics in arbitrary spacetime dimension $D$. The precise conditions for such chaos were presented in~\cite{Damour:2002et}, building on earlier contributions~\cite{Misner:1969ae,chitre1972,Ivashchuk:1994da,Ivashchuk:1994fg,Kirillov:1994fc}. In these works it was explained that the asymptotic evolution towards the singularity corresponds to geodesic motion in hyperbolic space, interrupted by abrupt bounces against the walls of a convex polyhedron, with chaos arising when the corresponding polyhedron has finite volume.

BKL dynamics has been explicitly observed in homogeneous cosmologies, both in vacuum gravity for $4\leq D\leq 10$ and in theories coupled to $p$-forms for general $D$~\cite{Misner1969,Belinsky:1970ew,1988PhLB..211...37D,Benini:2005su,Montani:2007vu}. However, explicit realizations of BKL dynamics in black hole interiors have remained largely elusive. Apart from its intrinsic relevance for gravitational physics, the study of the approach towards spacetime singularities is especially important in holographic contexts, as it can leave strong imprints on CFT observables, offering valuable insights into the meaning of bulk singularities from the boundary perspective~\cite{Fidkowski:2003nf,Festuccia:2005pi,Frenkel:2020ysx,Carballo:2024hem,Caceres:2022smh,Caceres:2023zhl,Rodriguez-Gomez:2021pfh,Dodelson:2023vrw,Jorstad:2023kmq,Arean:2024pzo,Horowitz:2023ury,Caceres:2023zft,Chakravarty:2025ncy,Ceplak:2024bja,Ceplak:2025dds}.
With this motivation in mind, several constructions exhibiting Kasner interiors with varied phenomenology have been developed in recent years~\cite{Hartnoll:2020rwq,Hartnoll:2020fhc,Sword:2021pfm,Sword:2022oyg,Wang:2020nkd,Mansoori:2021wxf,Liu:2021hap,Das:2021vjf,An:2022lvo,Auzzi:2022bfd,Mirjalali:2022wrg,Hartnoll:2022snh,Hartnoll:2022rdv,Caceres:2022hei,Liu:2022rsy,Gao:2023zbd,Blacker:2023ezy,Caceres:2024edr}. Although these do not display truly chaotic dynamics, the first explicit realization of genuinely BKL dynamics in a black hole interior was recently achieved in~\cite{DeClerck:2023fax}, where four-dimensional GR was coupled to three massive vector fields in AdS and it was shown that BKL dynamics emerges deep inside the black hole, near the spacelike singularity.

Understanding BKL dynamics in holography may be even more compelling than in cosmological applications because the number of epochs can be parametrically enhanced. Indeed, although BKL predicts an infinite sequence of Kasner periods, the cascade cannot continue indefinitely within classical GR: as the singularity is approached, curvatures grow without bound and eventually exceed the regime of validity of the classical description. A convenient way to quantify this is to introduce a time variable $T$ such that the typical interval between successive wall collisions is approximately constant~\cite{Damour:2002et}. As the proper time $t\to 0$, one finds
\begin{equation}
T \sim \ln\left|\ln(t/t_\ast)\right|\,,
\label{eq:Tdoublelog}
\end{equation}
where $t_\ast$ is a macroscopic reference scale (e.g.\ the Hubble time in cosmology, or the AdS radius in holographic setups) introduced to make the logarithms dimensionless. This implies that the number of Kasner transitions accumulated down to time $t$ grows double-logarithmically. Let $t_0$ be the (already small) proper time at which the BKL approximation becomes reliable, and let $t_{\rm qg}$ be the time at which the curvature reaches the quantum-gravity cutoff. Then the number of Kasner transitions that can be trusted within classical gravity scales as
\begin{equation}
N_{\rm epochs}\sim \mathcal{O}(1)\times
\left[
\ln\ln\!\left(\frac{t_\ast}{t_{\rm qg}}\right)
-\ln\ln\!\left(\frac{t_\ast}{t_0}\right)
\right].
\label{eq:Nepochs_general}
\end{equation}
For reasonable hierarchies in our universe, this typically yields only a handful of bounces before the classical description breaks down~\cite{Damour:2002et}. In particular, one may not be able to reach even a change of era while remaining safely within classical GR.

In AdS/CFT-like setups, however, classical bulk gravity corresponds to the large-$N$ limit: bulk quantum loops are suppressed by powers of $G_N$, mapping to $1/N$ (more precisely $1/N^2$) effects on the boundary theory. Taking $t_\ast\sim L$ (the AdS radius) and $t_{\rm qg}\sim \ell_p$, the hierarchy $L/\ell_p$ is set by $N$ via $L^{D-2}/G_N\propto N^2$. Consequently, the classical window accommodates
\begin{equation}
N_{\rm epochs}\sim \mathcal{O}(1)\times \ln\ln\!\left(\frac{L}{\ell_p}\right)
\sim \ln\ln N,
\label{eq:Nepochs_holo}
\end{equation}
epochs. Increasing $N$ therefore extends the number of classical epochs (albeit very slowly), potentially allowing access to a regime in which multiple eras are trustworthy. At any finite $N$, the late stages of the BKL cascade should, however, receive significant corrections from non-planar effects in the dual field theory, taking us outside classical GR.

Investigations of the approach towards spacelike singularities in holographic black hole interiors have continued~\cite{Cai:2023igv,Gao:2023rqc,Cai:2024ltu,DeClerck:2025mem,Caceres:2025xzl,Duan:2026qhj}. Nonetheless, explicit realizations remain scarce. While a five-dimensional example has been explored in~\cite{DeClerck:2025mem}, a detailed characterization of the resulting dynamics is still incomplete---for instance, the refined organization of eras (and the associated Kasner seasons identified here) was not uncovered there---and no explicit realization is known for $D>5$. The purpose of this paper is to provide such a construction. We will show that the model we present exhibits chaotic BKL dynamics deep in the black hole interior in arbitrary spacetime dimension $D\geq 4$. We will also study in detail the ensuing structure of Kasner eras in $D=5$ as a first step towards a more systematic characterization for $D>5$, and we will compare with the chaotic dynamics of purely gravitational five-dimensional homogeneous cosmologies. Finally, as a holographic application, we present a detailed study of the thermal $a$-function~\cite{Caceres:2022smh}, which we show is sensitive to the transitions between Kasner epochs and can therefore serve as a probe of the near-singularity structure from the CFT side.

The paper is organized as follows. In Section~\ref{sec:review_4d} we review aspects of the four-dimensional background exhibiting BKL dynamics. Section~\ref{sec:bkl_higher_d} contains our main results: we generalize the previous construction to higher dimensions $D\geq 5$, show the appearance of chaotic BKL dynamics sufficiently close to the singularity, derive the bouncing rules for the Kasner exponents between consecutive epochs, and identify a novel notion of \emph{Kasner seasons}, which serves to distinguish the various types of epochs within eras in $D\geq 5$. We present, along the way, a variety of numerical examples supporting our theoretical developments. Next, in Section~\ref{sec:holographic_probe} we elaborate on the thermal $a$-function as a possible probe of the black hole interior. We conclude with a brief summary of our results and discuss various future directions opened by our work in Section~\ref{sec:discussion}.







\section{BKL dynamics in four-dimensional AdS black holes: a review}\label{sec:review_4d}

\subsection{Bulk action and equations of motion}
We begin by reviewing the four-dimensional construction presented in \cite{DeClerck:2023fax}, which explicitly exhibits BKL dynamics inside a black hole interior. Consider \((3+1)\)-dimensional Einstein gravity minimally coupled to three massive vector fields \(\{A_i\}_{i=1}^3\) as follows:
\begin{equation}
    S=\frac{1}{16 \pi G_N} \int \mathrm{d}^{4} x \sqrt{ \vert g \vert} \left [R+\frac{6}{\ell^2}- \sum_{i=1}^{3} \left\lbrace \frac{F_i^2}{4}+ \frac{\mu_i^2}{2}A_i^2 \right\rbrace \right] \,,
\end{equation}
where \(\ell\) denotes the AdS radius and \(\mu_i^2\) the mass squared of the corresponding vector field. We assume the following ansatz for the metric and the vector fields:
\begin{align}
\label{eq:metricbh4}
    \mathrm{d}s^2&=\frac{1}{z^2} \left ( - F e^{-2H} \mathrm{d}t^2+ \frac{\mathrm{d}z^2}{F}+ e^{-\sqrt{3}G_1} \mathrm{d}x^2+ e^{\sqrt{3}G_1} \mathrm{d}y^2 \right)\,, \\
    A_1&=\phi_t \mathrm{d}t\,, \quad A_2=\phi_x \mathrm{d} x\,, \quad  A_3=\phi_y \mathrm{d} y\,,
\end{align}
where \(\{F,H,G_1,\phi_t,\phi_x,\phi_y\}\) are functions of \(z\). The solutions are asymptotically AdS, with boundary located at \(z=0\), and describe a black hole whenever \(F\) has a zero. In what follows, we assume this to be the case. Furthermore, we require one of the vector fields to have negative mass squared --- which is allowed by AdS asymptotics\footnote{Nonetheless, it should be noted that, although a negative bulk mass squared may satisfy the Breitenlohner--Freedman bound, it can still be holographically dual to a boundary operator whose scaling dimension lies below the CFT unitarity bound. We refer to \cite{DeClerck:2023fax} for further details.} as long as it remains above the Breitenlohner--Freedman stability bound \cite{Breitenlohner:1982jf,Breitenlohner:1982bm} --- since this guarantees that the black hole possesses a single horizon and, consequently, a spacelike singularity. Therefore, \(F\) has a single root \(z_h\) in the region \(z>0\), and \(F<0\) for all \(z>z_h\). The singularity arises as \(z \to \infty\).

In the black hole interior $F<0$, it is convenient to change the coordinate $z$ by a new (time-like) coordinate $\rho$ defined as follows:
\begin{equation}
    z^2=e^{\rho-f_0(\rho)}\,, \quad -F(z) e^{-2H(z)}=z^2 e^{-\rho-2 f_0(\rho)}\,, \quad n(\rho)^2 \mathrm{d}\rho^2=-\frac{\mathrm{d}z^2}{F(z)z^2}\,,
\end{equation}
where we have implicitly defined new functions $f_0(\rho)$ and $n(\rho)$ in terms of $F$ and $H$. Now, \eqref{eq:metricbh4} may be written as:
\begin{equation}
    \mathrm{d}s^2=-n(\rho)^2 \mathrm{d}\rho^2+e^{-\rho} e^{-2f_0} \mathrm{d}t^2+e^{-\rho} e^{f_0} \left[e^{-\sqrt{3}f_1} \mathrm{d}x^2+e^{\sqrt{3}f_1} \mathrm{d}y^2\right]\,,
    \label{eq:intbh4}
\end{equation}
where $f_1(\rho)=G_1(z(\rho))$. The volume of constant $\rho$ slices is precisely given by $e^{-3\rho}$, collapsing to zero as one approaches the singularity as $\rho \rightarrow \infty$. As it turns out, the equations of motion for the gauge fields can be analytically integrated once into:
\begin{equation}
\dot{\phi}_t=\sqrt{2\xi_0}\, n \, e^{\rho/2-2f_0}\,, \quad \dot{\phi}_x=\sqrt{2\xi_1}\, n \, e^{\rho/2-\sqrt{3} f_1+f_0}\,, \quad \dot{\phi}_y=\sqrt{2\xi_2}\, n \,  e^{\rho/2+\sqrt{3} f_1+f_0}\,,
\label{eq:pots4}
\end{equation}
where $\left\lbrace \sqrt{\xi_i} \right\rbrace_{i=1}^3$ are some real constants. Plugging these results into the Einstein equations, and disregarding the cosmological constant and the mass square terms --- which will be irrelevant in the deep black hole interior --- one finds the following set of ordinary differential equations:
\begin{align}
\label{eq:eom41}
    n^2 V e^{2\rho} -3 \left (1-\dot{f}_0^2-\dot{f}_1^2 \right)&=0\,, \\
    \label{eq:eom42}
    \ddot{f}_0+\frac{1}{2} \left (\frac{1}{V} \frac{\partial V}{\partial f_0}-\dot{f}_0 \right)\left (1-\dot{f}_0^2-\dot{f}_1^2 \right)&=0\,, \\
    \label{eq:eom43}
      \ddot{f}_1+\frac{1}{2} \left (\frac{1}{V} \frac{\partial V}{\partial f_1}-\dot{f}_1 \right)\left (1-\dot{f}_0^2-\dot{f}_1^2 \right)&=0\,, 
\end{align}
where the potential $V$ is made of three exponential blocks $\{V_i\}_{i=0}^{2}$:
\begin{align}
    V&=V_0+V_1+V_2\,, \\ V_0=\xi_0 e^{-2f_0}\,, &\quad V_1=\xi_1  e^{f_0-\sqrt{3}f_1}\,, \quad V_2=\xi_2  e^{f_0+\sqrt{3}f_1}\,.
    \label{eq:pot4}
\end{align}
Using \eqref{eq:eom41}, \eqref{eq:eom42} and \eqref{eq:eom43}, one may get the equation:
\begin{equation}
\frac{\dot{n}}{n}=-\frac{1}{2} \left (2+\dot{f}_0^2+\dot{f}_1^2 \right)\,.
\label{eq:edon4}
\end{equation}
If the first term in \eqref{eq:eom41}, $n^2 V e^{2\rho}$, is negligible, it is possible to prove that the space-time metric is described by the well known Kasner solution \cite{kasner1921geometrical}, saying correspondingly that the solution enters into a fixed Kasner regime. Indeed, in such a case, \eqref{eq:eom41}, \eqref{eq:eom42} and \eqref{eq:eom43} boil down to:
\begin{equation}
    \ddot{f}_0=\ddot{f}_1=0\,, \quad \dot{f}_0^2+\dot{f}_1^2=1\,.
    \label{eq:haciakas4}
\end{equation}
As a consequence,
\begin{equation}
\label{eq:conskas4}
    f_0=f_0^{(0)}+\lambda_0 \rho\,, \quad f_1=f_1^{(0)}+\lambda_1 \rho\,, \quad  \lambda_0^2+\lambda_1^2=1\,,
\end{equation}
for constants $\lambda_0$, $\lambda_1$, $f_0^{(0)}$ and $f_0^{(1)}$. We may set $f_0^{(0)}=f_0^{(1)}=0$ by shifting $\rho$ if necessary. Now one can define:
\begin{equation}
    p_0=\frac{1+2 \lambda_0}{3}\,,\quad p_1=\frac{1-\lambda_0+\sqrt{3}\lambda_1}{3}\,,\quad p_2=\frac{1-\lambda_0-\sqrt{3}\lambda_1}{3}\,,
    \label{eq:kas4}
\end{equation}
and these parameters satisfy, based on their definition and \eqref{eq:haciakas4}:
\begin{equation}
    p_0+p_1+p_2=1\,, \quad p_0^2+p_1^2+p_2^2=1\,.
\end{equation}
Furthermore, within a Kasner regime  \eqref{eq:edon4} implies that $n=n_0\, e^{-3\rho/2}$ for some constant $n_0$, so one can rewrite \eqref{eq:intbh4} as:
\begin{equation}
    \mathrm{d}s^2=-\frac{e^{-3\rho}}{n_0^2} \mathrm{d}\rho^2+ c_t^2 e^{-3\rho \, p_0} \mathrm{d}t^2+ c_x^2 e^{-3\rho \, p_1} \mathrm{d}x^2+ c_y^2 e^{-3\rho \, p_2} \mathrm{d}y^2\,,
    \label{eq:kasnerformmet}
\end{equation}
\begin{figure}
   \centering
\includegraphics[width=0.75\linewidth]{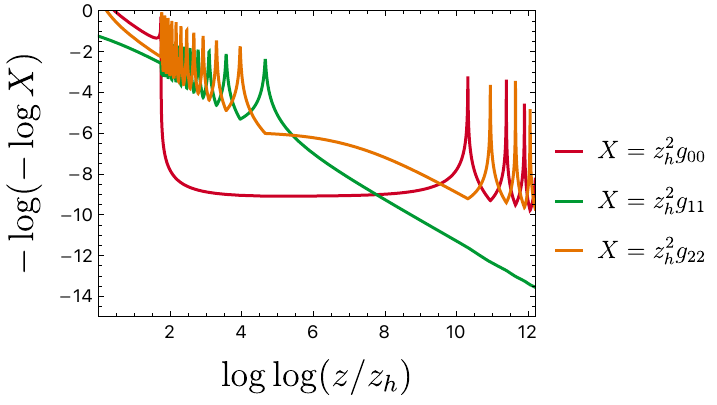}
   \caption{Appearance of BKL dynamics inside a four-dimensional AdS black hole. We show the interior-time evolution of the three independent diagonal components of the metric, $\displaystyle ds^{2}=g_{zz}\,dz^{2}+\sum_{i=0}^{2} g_{ii}\,(dx^{i})^{2}$.}
    \label{fig:bkl4d}
\end{figure}
for some innocent constants $c_t,c_x,c_y$. Defining a new coordinate $\tau \propto e^{-3\rho/2}$ appropriately, one may in fact write:
\begin{equation}
        \mathrm{d}s^2=-\mathrm{d}\tau^2 + \tau^{2p_0} \mathrm{d}t^2+ \tau^{2p_1} \mathrm{d}x^2+\tau^{2p_2}\mathrm{d}y^2\,,
    \label{eq:kasnerfourmet}
\end{equation}
where the spatial coordinates have also been rescaled. From here, it becomes evident that the regime in which the first term in \eqref{eq:eom41} is subleading corresponds to a Kasner solution with exponents \eqref{eq:kas4}. This behaviour is presented in Fig \ref{fig:bkl4d}. Here, we explicitly demonstrate through numerics that the late-time metric evolution exhibit oscillatory epochs grouped into eras--- we present two full Kasner eras composed of several Kasner epochs in this plot. 

Interestingly enough, if one defines the following \emph{effective} Kasner exponents:
\begin{equation}
    p_0^{\rm eff}=-\frac{1}{3} \partial_\rho \log g_{00}\,, \quad  p_1^{\rm eff}=-\frac{1}{3} \partial_\rho \log g_{11}\,, \quad 
    p_2^{\rm eff}=-\frac{1}{3} \partial_\rho \log g_{22}\,,
\end{equation}
in terms of the diagonal components of the metric \eqref{eq:intbh4}, Kasner exponents will correspond precisely to those regimes in which these effective exponents are constant --- and equal to the Kasner exponents defining the Kasner solution. This is illustrated in Figure \ref{4dtriangle} (left).

\subsection{Exponential walls and bouncing rules in $D=4$}

\begin{figure}
    \centering
    \includegraphics[width=0.49\linewidth]{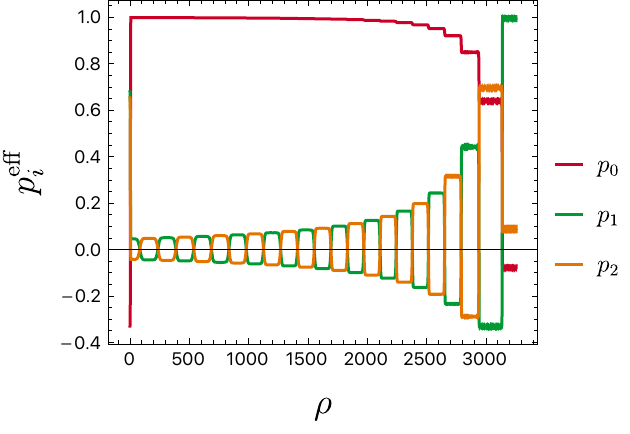}
    \includegraphics[width=0.48\linewidth]{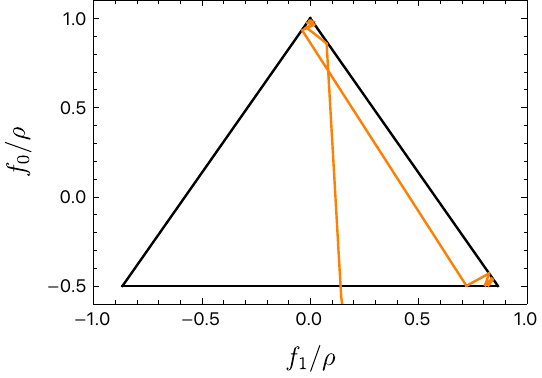}
    \caption{\textbf{Left:} Effective Kasner exponents in the black hole interior \eqref{eq:intbh4}. The different plateaus of the effective Kasner exponents correspond to Kasner epochs, which undergo very rapid transitions into new Kasner epochs. Observe the presence of a first (long) Kasner era, which is the result of some appropriate boundary conditions that produce this illustrative result --- other choices of boundary conditions are possible, giving rise to shorter initial eras. We notice that, around $\rho \sim 2900$ and $\rho \sim 3200$, the system undergoes changes of Kasner eras. \textbf{Right:}
    Approach to the singularity in the configuration space defined by $\left ( \frac{f_0}{\rho}, \frac{f_1}{\rho} \right)$. It is explicitly checked that it consists of straight lines (Kasner epochs) interrupted by abrupt bounces to the edges of an equilateral triangle. Kasner eras correspond to those collections of bounces between the same two edges. }
    \label{4dtriangle}
\end{figure}

The Kasner approximation will be valid as long as the first term in \eqref{eq:eom41} is subleading. This will be no longer true for values $\rho_{\rm e}$ of the coordinate $\rho$ such that:
\begin{equation}
    n^2 V_{\rm e} e^{2 \rho_{\rm e}} \sim 1\,,
\end{equation}
where $V_{\rm e}=V(\rho_{\rm e})$. Using that $n \propto e^{-3\rho/2}$, we find the condition:
\begin{equation}
   \rho_{\rm e}\sim \log V_{\rm e}\,.
\end{equation}
Now, being in the deep interior, $\rho_{\rm e}$ will be particularly large, with $V_{\rm e}$ being the exponential thereof. To leading exponential precision, we may express:
\begin{equation}
\log V_{\rm e}= \mathrm{max} \left\lbrace-2f_0(\rho_{\rm e}),f_0(\rho_{\rm e})-\sqrt{3}f_1(\rho_{\rm e}), f_0(\rho_{\rm e})+\sqrt{3}f_1(\rho_{\rm e}) \right\rbrace
\end{equation}
As a result, sufficiently close to the black hole singularity, by virtue of \eqref{eq:conskas4} we may approximate the potential to be an infinite potential well $V_\infty$ in the parameter space $\left ( \frac{f_0}{\rho}, \frac{f_1}{\rho} \right)$ of the form:
\begin{equation}
\label{eq:walls4pot}
    V_\infty=\begin{cases} 
      0 & \dfrac{f_0}{\rho} \geq -\dfrac{1}{2}\,, \quad \dfrac{f_0}{\rho} \pm\sqrt{3}\, \dfrac{f_1}{\rho} \leq 1  \\
      \infty & \text{otherwise}
   \end{cases}
\end{equation}
Beyond these bounds, the exponential walls are assumed to be sufficiently steep to be regarded as infinite walls to a high degree of approximation. The region in which $V_\infty=0$ conforms an equilateral triangle. Its edges are given by walls $\left\lbrace\mathrm{W}_m \right\rbrace_{m=0}^2$:
\begin{equation}
\label{eq:walls4}
    \left\lbrace {\rm Wall}\, \, {\rm W}_0: \frac{f_0}{\rho}=\frac{1}{2}\,,  \quad {\rm Wall}\, \, {\rm W}_1: \frac{f_0-\sqrt{3}f_1}{\rho}=1\,, \quad  {\rm Wall}\, \, {\rm W}_2: \frac{f_0+\sqrt{3}f_1}{\rho}=1 \right\rbrace\,,
\end{equation}
where the wall $\mathrm{W}_m$ is associated to the exponential block $V_m$ in \eqref{eq:pot4} for $m=0,1,2$. As a result, the dynamical evolution of the system may be realized as the motion of a particle in the interior of an equilateral triangle, following a straight line which ends as it reaches any of the edges of the equilateral triangle. At this point, a \emph{bounce}\footnote{Note that the bounces against the walls do not represent exact reflections. As a matter of fact, the incident and reflected angles are not the same in general. This is to be compared to the picture of hyperbolic billiards \cite{Ivashchuk:1994fg,Kirillov:1994fc,Damour:2002et}, where one conceives dynamics as geodesic motion in hyperbolic space interrupted by hyperbolic reflections. We would like to remark that such a realization is equivalent to the one we are presenting here, where we are resorting to a different configuration space in which motion occurs as free motion in Euclidean space constrained to the interior of an equilateral triangle.} occurs and the particle continues its motion within the equilateral triangle, pursuing a straight line with a different slope --- see Figure \ref{4dtriangle}. Each of these straight lines corresponds to regimes 
in which the metric may be locally approximated by a Kasner solution determined by some fixed Kasner exponents: \emph{Kasner epochs}. After a bounce, Kasner exponents suffer an abrupt change and the system transitions into a new Kasner epoch with different (and well defined) Kasner exponents. 

Let us now relate the motion in the parameter space $\left(\frac{f_0}{\rho},\frac{f_1}{\rho} \right)$ to the usual picture of hyperbolic billiards \cite{Ivashchuk:1994fg,Kirillov:1994fc,Damour:2002et}. Following the seminal work \cite{Damour:2002et}, the \emph{logarithmic scale factors}  $\{\beta_i\}_{i=1}^3$ (defined as $g_{ii}=e^{-2\beta_i}$) correspond to: 
\begin{equation}
    \beta_1=\frac{\rho}{2} +f_0(\rho)\,, \quad \beta_2=\frac{\rho}{2} -\frac{f_0(\rho)}{2}+\frac{\sqrt{3}}{2}f_1(\rho)\,, \quad \beta_3=\frac{\rho}{2} -\frac{f_0(\rho)}{2}-\frac{\sqrt{3}}{2}f_1(\rho)\,. 
\end{equation}
The relevant walls which will constrain the motion on hyperbolic space (after decomposing the $\{\beta_i\}_{i=1}^3$ into radial and angular components) are of electric origin, arising from the three vector fields. By direct comparison to \eqref{eq:pot4}, after using that $n=n_0 e^{-3\rho/2}$ during a Kasner epoch, it is clear that motion is constrained to the following region of $\beta$-space:
\begin{equation}
    \beta_1 \geq 0\,, \quad \beta_2 \geq 0\,, \quad \beta_3 \geq 0\,.
\end{equation}
We observe that this coincides with the electric walls considered in \cite{Damour:2002et}. These form a finite-volume hyperbolic billiard and result in chaotic dynamics.


Now, let us briefly derive the \emph{bouncing rules} or \emph{collision laws} that relate Kasner exponents before and after a certain bounce on a given edge of the triangle. To this aim, let us examine the full equations of motion \eqref{eq:eom42} and \eqref{eq:eom43}. By dividing them, we find:
\begin{equation}
    \frac{\ddot{f}_1}{\ddot{f}_0}=\frac{\frac{1}{V} \frac{\partial V}{\partial f_1}-\dot{f}_1}{\frac{1}{V} \frac{\partial V}{\partial f_0}-\dot{f}_0}\,.
    \label{eq:ddot4}
\end{equation}
Remember from \eqref{eq:pot4} that the potential $V$ was built from three exponential blocks giving rise to the walls identified in \eqref{eq:walls4}. In a bounce, $f_0$ and $f_1$ remain constant, while their derivatives change abruptly. Consequently, we may take the potential $V$ and their derivatives to be constant during the transition between consecutive Kasner epochs. Define:
\begin{equation}
    \lambda_0^{(i)}=\left. \frac{1}{V} \frac{\partial V}{\partial f_0} \right \vert_{{\rm W}_m}\,, \quad \lambda_1^{(i)}=\left. \frac{1}{V} \frac{\partial V}{\partial f_1} \right \vert_{{\rm W}_m}\,, \quad m=0,1,2\,,
\end{equation}
where $\vert_{{\rm{W}}_m}$ stands for evaluation in the $m$-th wall given in \eqref{eq:walls4}. Direct computation shows:
\begin{align}
    {\rm W}_0: \quad  \lambda_0^{(0)}&=-2\,, \quad \lambda_1^{(0)}=0\,, \\
     {\rm W}_1: \quad  \lambda_0^{(1)}&=1\,, \quad \lambda_1^{(1)}=-\sqrt{3}\,, \\
      {\rm W}_2:  \quad \lambda_0^{(2)}&=1\,, \quad \lambda_1^{(2)}=\sqrt{3}\,.
\end{align}
Considering \eqref{eq:ddot4} in a region $\left(\frac{f_0}{\rho},\frac{f_1}{\rho} \right)$ sufficiently close to a wall ${\rm W}_m$, we may take $\left\lbrace \left. \dfrac{1}{V} \dfrac{\partial V}{\partial f_0} \right \vert_{{\rm W}_m},\left. \dfrac{1}{V} \dfrac{\partial V}{\partial f_1} \right \vert_{{\rm W}_m}\right\rbrace_{m=0}^{2}$ to be the constants $\left\lbrace \lambda_0^{(m)},\lambda_1^{(m)}\right\rbrace_{m=0}^{2}$ to a high degree of approximation and integrate \eqref{eq:ddot4} to obtain the first-order equation:
\begin{equation}
    \frac{\dot{f}_1-\lambda_1^{(m)}}{\dot{f}_1^{(0)}-\lambda_1^{(m)}}= \frac{\dot{f}_0-\lambda_0^{(m)}}{\dot{f}_0^{(0)}-\lambda_0^{(m)}}\,, \quad m=0,1,2\,,
    \label{eq:eqparkas4}
\end{equation}
where $\left ( \dot{f}_0^{(0)}, \dot{f}_1^{(0)} \right)$ denote the values before the bounce and $\left ( \dot{f}_0, \dot{f}_1 \right)$ after it. Interestingly enough, \eqref{eq:eqparkas4} may be rewritten in an extremely simple form in terms of the exponents $\left ( p_0^{(0)}, p_1^{(0)},p_2^{(0)} \right)$ of the Kasner epoch before the bounce and the Kasner exponents $\left ( p_0, p_1,p_2 \right)$ of the successive Kasner epoch, arising after the bounce onto the corresponding wall. Specifically, after some algebraic work one may prove analytically that \eqref{eq:eqparkas4} is equivalent to:
\begin{equation}
    {\rm W}_m: \quad  \, \frac{p_i}{p_{m}}=-2-\frac{p_i^{(0)}}{p_{m}^{(0)}}\,, \quad i,m=0,1,2 \,, \quad i\neq {m}\,.
\end{equation}
Together with the condition $p_0+p_1+p_2=1$, we get a linear system of equations for $(p_0,p_1,p_2)$. Its solution is given by:
\begin{equation}
    p_i=\frac{p_i^{(0)}+2p_{m}^{(0)}}{1+2p_{m}^{(0)}}\,, \quad p_{i}=-\frac{p_{m}^{(0)}}{1+2p_{m}^{(0)}}\,, \quad i=0,1,2 \,, \quad i\neq {m}\,.
\end{equation}
This is precisely the bouncing rule that one gets for the standard four-dimensional BKL or Mixmaster model \cite{Belinsky:1970ew}. It can be simplified even further if one introduces the following one-variable parametrization:
\begin{equation}
    \mathfrak{p}_2(u)=\frac{u(1+u)}{1+u+u^2}\,,\quad  \mathfrak{p}_1(u)=\frac{1+u}{1+u+u^2}\,, \quad \mathfrak{p}_0(u)=\frac{-u}{1+u+u^2}\,,
\end{equation}
where the real variable $u \geq 1$ and $1 \geq \mathfrak{p}_2(u) \geq \dfrac{2}{3} \geq \mathfrak{p}_1(u) \geq 0 \geq \mathfrak{p}_0(u) \geq -\dfrac{1}{3}$. If the initial exponents satisfy $p_2^{(0)}=\mathfrak{p}_2(u)$, $p_1^{(0)}=\mathfrak{p}_1(u)$ and $p_0^{(0)}=\mathfrak{p}_0(u)$, then the collision law for exponents are fully characterized in by the following rules \cite{Misner1969,Belinsky:1970ew,lifshitz1971asymptotic}:
\begin{enumerate}
    \item If $u \geq 2$:
    \begin{equation}
    \label{eq:col1}
        p_2=\mathfrak{p}_2(u-1)\,, \quad   p_1=\mathfrak{p}_0(u-1)\,, \quad   p_0=\mathfrak{p}_1(u-1)\,.
    \end{equation}
    In particular, the new exponents are ordered as $p_2 \geq p_0 \geq p_1$\,.
    \item If $2 \geq u \geq 1$:
    \begin{equation}
    \label{eq:col2}
        p_2=\mathfrak{p}_1\left (\frac{1}{u-1}\right )\,, \quad   p_1=\mathfrak{p}_0\left (\frac{1}{u-1}\right )\,, \quad   p_0=\mathfrak{p}_2\left (\frac{1}{u-1}\right )\,.
    \end{equation}
    Now, the new exponents are ordered as $p_0 \geq p_2 \geq p_1$\,.
\end{enumerate}
Let us comment on the nature of the two transition rules. As may be readily seen from \eqref{eq:kasnerformmet}, positive Kasner exponents represent contracting directions, while negative ones are associated with expanding directions. In four dimensions, there will be two contracting directions and an expanding one. Consequently, the first type of collision law \eqref{eq:col1}
preserves the largest direction of contraction while exchanging the character of the other two directions.  Sequences of consecutive Kasner epochs followed by bounces of this type, which preserve the direction of largest contraction rate, conform \emph{Kasner eras}. On the other hand, the second law \eqref{eq:col2} explicitly changes the direction of the largest exponent, thus being interpreted as 
a change of Kasner era. The subsequent BKL dynamics are known to be chaotic and we refer the reader to \cite{Misner1969,Belinsky:1970ew,Damour:2002et,Belinski:2017fas} for additional information on the topic.

\section{BKL dynamics in higher-dimensional AdS black holes}\label{sec:bkl_higher_d}

\subsection{Bulk action and equations of motion}

The purpose of this section is to generalize the construction presented in \cite{DeClerck:2023fax} to generic space-time dimensions. To this aim, let us consider the following $D$-dimensional theory of gravity with $(D-1)$ massive gauge fields\footnote{Note that the auxiliary length scale $\ell$ is related to the AdS radius $L_{\text{AdS}}$ via $L_{\text{AdS}}=\ell\,\sqrt{\frac{2(D-2)}{D}}\,$. In the numerical analysis we fix the overall length scale by setting $\ell=1$. This choice is made purely for numerical convenience: it improves the conditioning and stability of the boundary value problem.
Physical observables can be restored to arbitrary $\ell$ (and hence arbitrary $L_{\text{AdS}}$) by dimensional analysis.}:
\begin{equation}
    S=\frac{1}{16 \pi G_N} \int \mathrm{d}^{D} x \sqrt{ \vert g \vert} \left [R+\frac{D(D-1)}{2\ell^2}-  \sum_{i=0}^{D-2}\left\lbrace \frac{F_i^2}{4}+ \frac{\mu_i^2}{2}A_i^2\right\rbrace \right] \,.
\end{equation}
The equations of motion are given by:
\begin{align}\label{eq: Einsteinplusmatter}
    R_{ab}-\frac{R}{2}g_{ab}-\frac{D(D-1)}{4 \ell^2} g_{ab}&=\frac{1}{2}\sum_{i=0}^{D-2} \left (F_i{}_{ac} F_i{}_{b}{}^c-\frac{g_{ab}}{4} F_i^2+ \mu_i^2 \left[ A_{i}{}_{a}A_{i}{}_{b}-\frac{g_{ab}}{2} A_i^2 \right] \right)\,, \\
    \label{eq:eqvec}
    \nabla_a F_i^{ab}-\mu_i^2 A_i^b&=0\,, \quad \quad\, \qquad i=0,\dots, D-2\,.
\end{align}
We stress that, all along the document, repeated Latin indices will not imply summation over them, unless otherwise specified. Let us consider the following ansatz for the metric and vector fields:
\begin{align}
\label{eq:metricbh}
    \mathrm{d}s^2&=\frac{1}{z^2} \left ( - F e^{-2H} \mathrm{d}t^2+ \frac{\mathrm{d}z^2}{F}+ \sum_{p=1}^{D-2} e^{-(D-2-p) a_p f_p } \left ( \prod_{q=1}^{p-1} e^{a_q f_q} \right)  \left (\mathrm{d}x^{p} \right)^2 \right)\,, \\
    A_0&=\phi_0(z) \mathrm{d}t\,, \quad A_{p}=\phi_{p} (z) \mathrm{d} x^{p}\,, \quad p=1,\dots, D-2 
\end{align}
where $F$,$H$ and $\{f_p \}_{p=1}^{D-3}$ are functions of $z$, the coordinates $\{x^p\}_{p=1}^{D-2}$ are spatial and the constants $\{a_j\}_{j=0}^{D-3}$ are defined by\footnote{Observe that $f_{D-2}(\rho)$ and $a_{D-2}$ are not defined, as these will be automatically absent from \eqref{eq:metricbh}.}:
\begin{equation}
    a_j=\sqrt{\frac{(D-1)(D-2)}{(D-1-j)(D-2-j)}}\,, \quad j=0,\dots, D-3\,.
\end{equation}
These parameters satisfy the following relations:
\begin{equation}
    \sum_{k=0}^{j-1} a_k^2+(D-2-j)^2 a_j^2=(D-2)^2\,, \quad    \sum_{k=0}^{j-1} a_k^2-(D-2-j) a_j^2=-(D-2)\,, \quad  j=0,\dots, D-3\,.
    \label{eq:recparam}
\end{equation}
Observe that $\sqrt{\vert g \vert}=\frac{e^{-H}}{z^D}$. It is useful to obtain the expression for the equations for the potentials $\{ \phi_p \}_{p=0}^{D-2}$ from \eqref{eq:eqvec}. These read:
\begin{equation}
    \frac{\mathrm{d}}{\mathrm{d}z} \left ( \frac{e^{-H}F}{z^{D-2}} g^{ii} \phi'_i\right)=\mu_i^2 \phi_i e^{-H} z^{-D} g^{ii}\,, \quad i=0,\dots, D-2\,,
    \label{eq:maxuso}
\end{equation}
where $\phi_i'=\frac{\mathrm{d} \phi_i}{\mathrm{d}z}$. Now, metric \eqref{eq:metricbh} will correspond to a black hole space-time if $F$ presents zeros. In such a case, if at least one of the $\mu_p^2$ for $D-2 \geq p \geq 1$ is negative\footnote{We stress that this makes sense from the bulk perspective, although the holographic description would correspond to a non-unitary operator.} --- but above the Breitenlohner-Freedman bound ---,  we are going to show that the event horizon cannot be accompanied by inner horizons. Specifically, assume there are both inner and outer horizons at $z_O$ and $z_I$ respectively (i.e., $F(z_I)=F(z_O)=0$). In the region determined by $F<0$, direct use of \eqref{eq:maxuso} reveals:
\begin{equation}
    0=\int_{z_O}^{z_I} \frac{\mathrm{d}}{\mathrm{d}z} \left ( \frac{e^{-H}F}{z^{D-2}} g^{pp} \phi'_p \phi_p\right) \mathrm{d}z=\int_{z_O}^{z_I} g^{pp} e^{-H} \left (\frac{\mu_p^2 \phi_p^2}{z^{D}}+\frac{F (\phi_p')^2}{z^{D-2}} \right) \mathrm{d}z\,.
\end{equation}
Since $F<0$ and $\mu_p^2<0$, neglecting the case of identically vanishing potential, we conclude that an inner horizon $z_I$ may not occur. This argumentation generalizes that of \cite{Hartnoll:2020rwq} for higher space-time dimensions.

\begin{figure}[t!]
\centering
\begin{minipage}{0.49\linewidth}
    \includegraphics[scale=0.7]{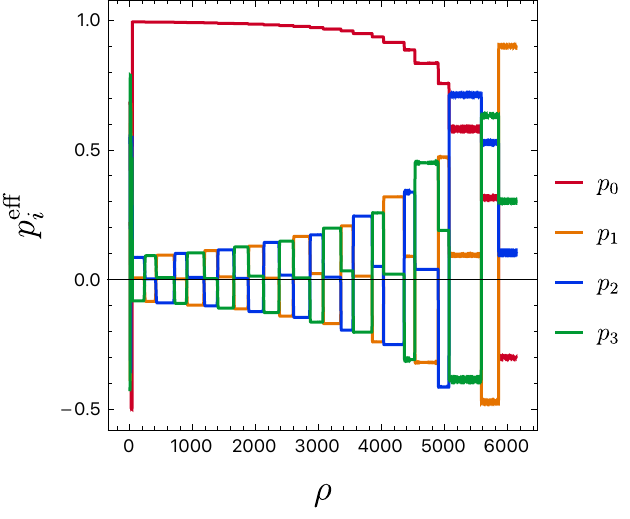}
\end{minipage}
\begin{minipage}{0.49\linewidth}
    \includegraphics[scale=0.7]{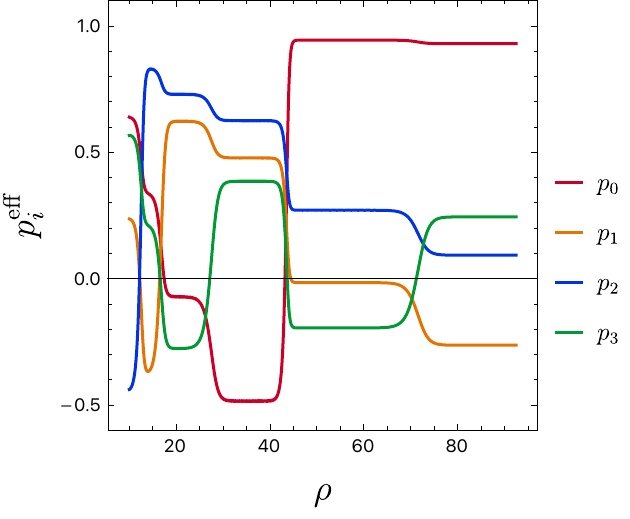}
\end{minipage}
\caption{Effective Kasner exponents for two different five-dimensional black holes fitting into the ansatz \eqref{eq:metricbh}. \textbf{Left:} We observe three complete Kasner eras. The last two are quite short and only contain one epoch, while the first one is quite long and features 20 epochs in Kasner season I --- see Definition \ref{def:ks}. \textbf{Right:} We see one complete era and a transition into a new one. We explicitly identify a Kasner epoch in Kasner season II. }
\label{fig:effKasnerExponents5}
\end{figure}

Having proven that there is a single horizon, let us study the approach to the singularity in the black hole interior. Since $F<0$ in this region, it is convenient to introduce a new coordinate $\rho$ and define new functions $f_0(\rho)$ and $n(\rho)$ as follows:
\begin{equation}
    z^2=e^{\rho-f_0(\rho)}\,, \quad -F(z) e^{-2H(z)}=z^2 e^{-\rho-(D-2) f_0(\rho)}\,, \quad n(\rho)^2 \mathrm{d}\rho^2=-\frac{\mathrm{d}z^2}{F(z) z^2}\,.
\end{equation}
Solving the first two equations, we find that:
\begin{equation}\label{domainwallcoord}
    \rho=-\frac{1}{D-1}\log\bigg(-\frac{e^{-2H(z)}F(z)}{z^{2(D-1)}}\bigg),\,\,\quad\,\, f_0(\rho)=-\frac{1}{D-1}\log(-e^{-2H(z)}F(z))
\end{equation}
We assume $n(\rho)$ is non-vanishing in the relevant coordinate region of $\rho$, so that it properly accounts for the black hole interior. Denoting as $\{f_p(\rho)\}_{p=1}^{D-2}$ the functions $\{f_p(z(\rho))\}_{p=1}^{D-2}$, we end up with the following coordinate expression for the metric and the gauge fields:
\begin{equation}
\label{eq:ans}
    \mathrm{d}s^2=-n(\rho)^2 \mathrm{d}\rho^2+\sum_{i=0}^{D-2} e^{-\rho} \frac{\partial V}{\partial \xi_i} \left (\mathrm{d}x^i \right)^2\,, \quad A_i=\phi_i (\rho) \mathrm{d} x^i\,, \quad i=0,\dots, D-2\,.
\end{equation}
where the coordinate $x^0=t$ and where we have defined\footnote{Following usual conventions, $\sum_{i=k}^{k-1} \sigma_i=0$ and $\prod_{i=k}^{k-1} \sigma_i=1$, for any argument $\sigma_i$ and integer $k$.  }:
\begin{align}
\label{eq:pot}
    V=\sum_{i=0}^{D-2} \xi_i\,  \mathrm{Exp} \left[ -(D-2-i)a_i f_i(\rho) +\sum_{j=0}^{i-1} a_j f_j(\rho) \right]  \,,
\end{align}
with $\{ f_i(\rho) \}_{i=0}^{D-3}$ being unknown functions of $\rho$ and $\{\xi_i\}_{i=0}^{D-2}$ positive constants. In the deep black hole interior, the masses of the vector fields do not play a relevant role and may be safely neglected. In such a regime, the subsequent Maxwell equations for $\{A_i\}_{i=0}^{D-2}$ when evaluated on \eqref{eq:ans} may be integrated into\footnote{Identifying the constants appearing in the potential $V$ with the charges in the potentials will highly simplify the notation, as it will be evident shortly.}
\begin{equation}
    \dot{\phi}_i=\sqrt{2\xi_i} \frac{\partial V}{\partial \xi_i} n(\rho) e^{(D-3) \rho/2}\,,  \quad i=0,\dots, D-2\,,
    \label{eq:potsol}
\end{equation}
where\footnote{Since the action is quadratic in the gauge fields, $\dot{\phi}_i=-\sqrt{2\xi_i} \frac{\partial V}{\partial \xi_i} n(\rho) e^{-(D-3) \rho/2}$ also produces a solution of the Maxwell equations. For our purposes, it will be indifferent whether we use $+ \sqrt{2\xi_i}$ or $- \sqrt{2\xi_i}$.} \emph{dot} stands for the derivative with respect to the coordinate $\rho$. Using this expression in the Einstein equations, it turns out that the whole set of Einstein equations boils down to the resolution of the following system of ordinary differential equations:
\begin{align}
\label{eq:eom1}
   & n^2 V e^{(D-2) \rho} -(D-1)\left (1-\sum_{i=0}^{D-3} \dot{f}_i^2\right) =0\,, \\ 
   \label{eq:eom2} &\ddot{f_j}+\frac{1}{2} \left ( \frac{1}{V}\frac{\partial V}{\partial f_j}-\dot{f_j} \right)\left (1-\sum_{i=0}^{D-3}    \dot{f}_i^2\right)=0\,, \quad j=0,\dots, D-3\,.
\end{align}
As in the four-dimensional case, whenever the piece $ n^2 V e^{(D-2) \rho}$ is subleading, the metric will be isometric to the higher-dimensional Kasner solution, giving rise to Kasner regimes. To prove this, assume that the first term in \eqref{eq:eom1} is negligible. Then, the equations of motion reduce to:
\begin{equation}
    \sum_{i=0}^{D-3} \dot{f}_i^2=1\,, \quad \ddot{f}_j=0\,, \quad j=0,\dots,D-3\,.
    \label{eq:sumlig}
\end{equation}
The solution to the previous equations is:
\begin{equation}
    f_j= f_j^0 +v_j \rho\,, \quad j=0,\dots,D-3\,, \quad \sum_{j=0}^{D-3} v_j^2=1\,,
    \label{eq:deff}
\end{equation}
where $\{f_j^0,v_j\}_{j=0}^{D-3}$ are integration constants. In fact, observe that if we define:
\begin{equation}
\label{eq:kasnerxp}
    p_i=\frac{1+(D-2-i)a_i v_i-\sum_{j=0}^{i-1} a_j v_j}{D-1}\,, \quad i=0,\dots, D-2\,,
\end{equation}
then, upon use of \eqref{eq:recparam} and \eqref{eq:deff}, 
\begin{equation}
\label{eq:kasnercons}
    \sum_{i=0}^{D-2} p_i=\sum_{i=0}^{D-2} p_i^2=1\,,
\end{equation}
which is formally equivalent to the constraints satisfied by the Kasner exponents in the $D$-dimensional Kasner solution. To see that these correspond to proper Kasner exponents, let us use the full equations of motion to obtain the following differential equation for the lapse $n(\rho)$:
\begin{equation}
    \frac{\dot{n}}{n}=-\frac{1}{2} \left (D-2+\sum_{i=0}^{D-3} \dot{f}_i^2 \right) \,.
\end{equation}
Under the assumption that the piece $n^2 V e^{(D-2)\rho}$ is subleading, one can integrate the previous equation and get $n \propto e^{-(D-1) \rho /2} $. Then, the metric adopts the following expression:
\begin{equation}
    \mathrm{d}s^2=-e^{-(D-1) \rho}\frac{\mathrm{d} \rho^2}{n_0^2}+\sum_{i=0}^{D-2} e^{-(D-1)\rho\,  p_i} \frac{\left ( \mathrm{d}x^i \right)^2}{c_i^2}
\end{equation}
for some innocent constants $n_0$ and $\left\lbrace c_i \right\rbrace_{i=0}^{D-2}$. Introducing a new coordinate $\tau \propto e^{-\frac{1}{2} (D-1) \rho}$,
we find
\begin{equation}
 \mathrm{d}s^2=-\mathrm{d}\tau^2+\sum_{i=0}^{D-2}\tau^{2p_i}(\mathrm{d}x^{i})^2\,,
\end{equation}
which corresponds to the $D$-dimensional Kasner space-time. Define the effective exponents 
\begin{equation}
    p_i^{\rm eff}=-\frac{1}{D-1}\partial_\rho \log g_{ii}\,, \quad i=0,\dots, D-2\,,
\end{equation}
as the logarithmic derivatives of the diagonal components of the metric \eqref{eq:ans}. If one tracks the evolution of these effective exponents towards the singularity, one would observe a sequence of \emph{plateaus} interrupted by rapid transitions --- see Figure \ref{fig:effKasnerExponents5}. Each regime of constant effective exponents conforms a different Kasner solution, while the sharp transitions occur whenever the first term in \eqref{eq:eom1} is no longer negligible, their net effect being the change of the specific Kasner solution approximating, locally, the metric. We will explore this aspect next.

\subsection{Exponential walls and bouncing rules in $D\geq4$}

\subsubsection{Motion constrained to a $(D-2)$-simplex}

As already mentioned, Kasner regimes take place as long as the term $n^2 V e^{(D-2) \rho} $  in \eqref{eq:eom1} is negligible. When it is no longer subleading, the Kasner regime breaks down and a transition into a new Kasner regime occurs, just like in the four-dimensional case. To see this, observe that during a Kasner regime  $n \propto e^{-(D-1) \rho /2} $, so that the first term in \eqref{eq:eom1} will become relevant for those values $\rho_n$ of the radial coordinate which satisfy:
\begin{equation}
    V_n e^{-\rho_n} \sim 1 \,, \longrightarrow \rho_n= \log V_n\,.
\end{equation}
From \eqref{eq:pot}, one notes that the potential $V$ is conformed by $D-1$ exponential blocks. To leading exponential accuracy, we may approximate $V_n$ as follows:
\begin{equation}
\label{eq:walls}
    \log V_n= \mathrm{max}\left\lbrace-(D-2-i) a_i f_i(\rho_n)+\sum_{j=0}^{i-1}a_j f_j(\rho_n)\,, \quad  i=0,\dots,D-3\right\rbrace\,.
\end{equation}
Consequently, as one gets sufficiently close to the singularity, $V$ adopts the form of an infinite potential well $V_\infty$ in the configuration space $\left\lbrace \dfrac{f_i}{\rho} \right\rbrace_{i=0}^{D-3}$:
\begin{equation}
  V_\infty=\begin{cases} 
      0 & \dfrac{(D-2-i)a_i f_i-\sum_{j=0}^{i-1} a_j f_j}{\rho} \geq -1\,, \quad \forall i=0,\dots,D-2  \\
      \infty & \text{otherwise}
   \end{cases}
\end{equation}
The region $V_\infty=0$ defines a $(D-2)$-dimensional polytope whose edges or walls $W_m$ are given by:
\begin{equation}
    \text{Wall  } {\rm{W}}_m: -(D-2-i)a_m\frac{f_m}{\rho} +\sum_{q=0}^{m-1} a_q \frac{f_q}{\rho} =1\,, \quad m=0,\dots, D-2\,.
    \label{eq:dwalls}
\end{equation}

Each wall $W_m$ conforms a hyperplane in the configuration space  $\left\lbrace \dfrac{f_m}{\rho} \right\rbrace_{m=0}^{D-3}$ whose unit normal vector $N_m$ is given by:
\begin{equation}
    N_m=\frac{1}{D-2}\left\lbrace a_0,a_1,\dots, a_{m-1},-(D-2-m)a_{m},0\dots,0 \right\rbrace\,, \quad m=0,\dots, D-2\,.
\end{equation}
Relations \eqref{eq:recparam} ensure that each $N_i$ is normalized to one and that:
\begin{equation}
    N_m \cdot N_q=-\frac{1}{D-2}\,, \quad \forall m>q\,,
\end{equation}

where ``$\cdot$'' stands for the usual Euclidean dot product. Amusingly, we notice that the (dihedral) angle between walls is given by $\mathrm{arccos} (\frac{1}{D-2} ) $, which is the defining feature of regular $(D-2)$-dimensional simplices. Therefore, motion is confined to the interior of a regular simplex: in $D=4$ within an equilateral triangles, in $D=5$ within a tetrahedron... Direct computation shows that its edges have (finite) length $\ell=\sqrt{\dfrac{2(D-1)}{(D-2)}}$ and the volume of the simplex is then given by $ \sqrt{\dfrac{D-1}{2^{D-2}}} \dfrac{\ell^{D-2}}{(D-2)!}$.

Furthermore, assuming that the Kasner regime approximation holds until one gets extremely close to the wall, one notes from the expression \eqref{eq:walls} that the free motion in the configuration space will get interrupted by collisioning with the wall associated with the smallest Kasner exponent \eqref{eq:kasnerxp} --- i.e., the most negative one.


Just like in four dimensions, the dynamical evolution of the system takes place as a sequence of free trajectories, confined to a regular simplex in the configuration space $\left\lbrace \dfrac{f_j}{\rho} \right\rbrace_{j=0}^{D-3}$, which are interrupted as these hit the walls defined by the regular (and compact) $(D-2)$-dimensional simplex. Whenever this happens, the initial Kasner regime will abruptly change into a new one, recovering the notion of Kasner epochs. As before, bounces against the faces of the simplex do not represent exact reflections, as the incident and reflected angles need not to be the same. In the usual hyperbolic billiard picture, these bounces would correspond to proper hyperbolic reflections. 

\begin{figure}[t!]
\centering
\begin{minipage}{0.49\linewidth}
    \includegraphics[scale=0.32]{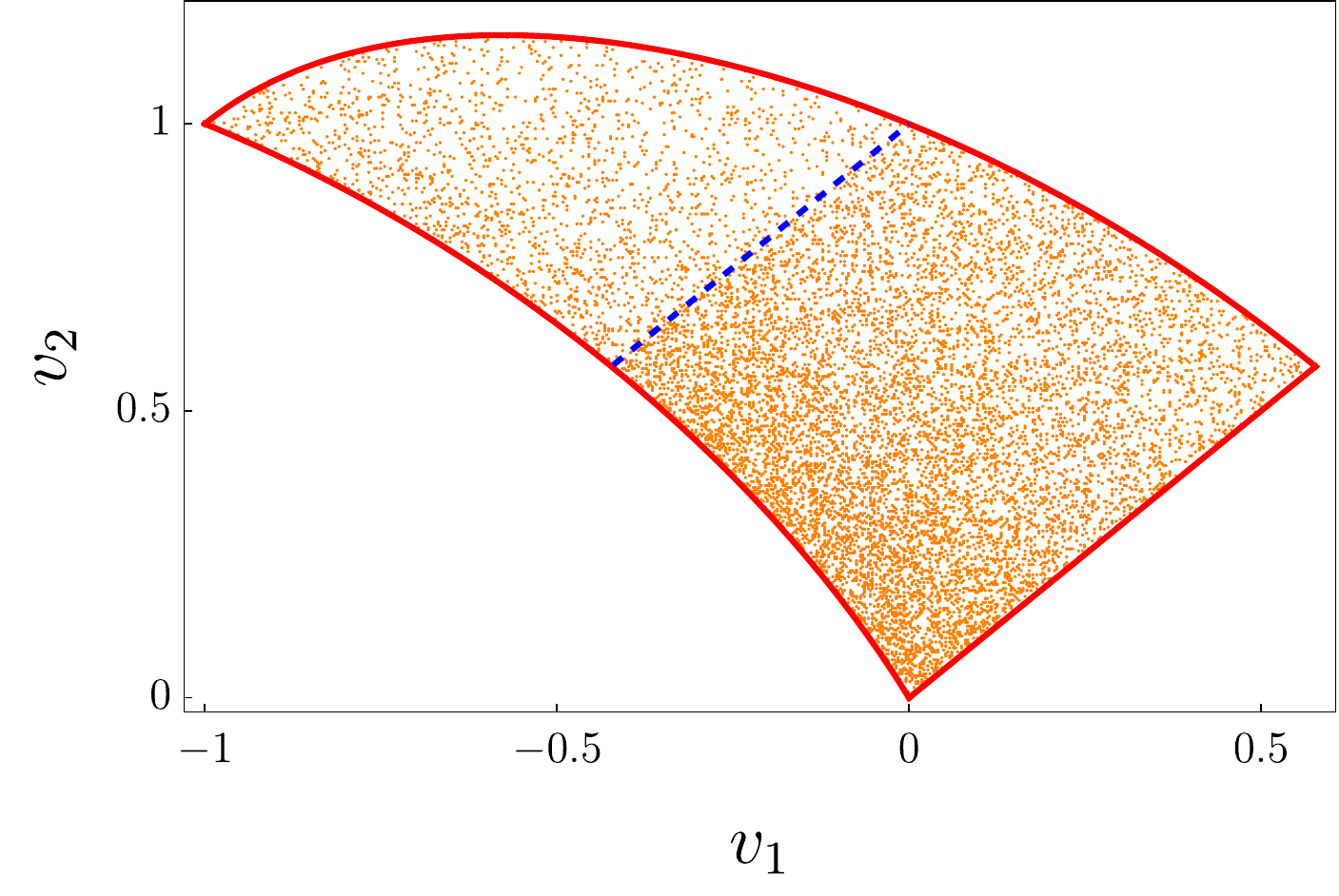}
\end{minipage}
\begin{minipage}{0.49\linewidth}
    \includegraphics[scale=0.32]{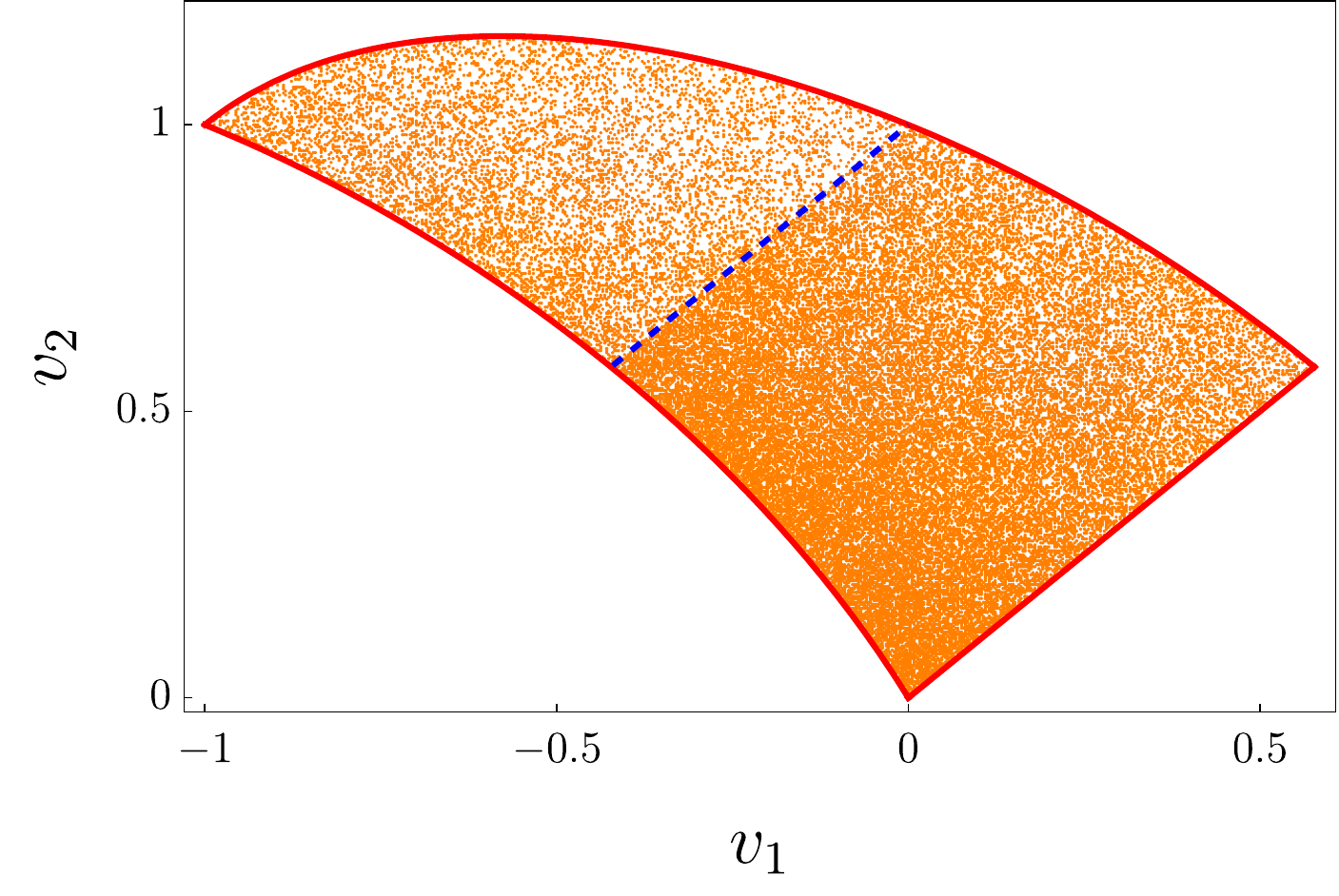}
\end{minipage}
\caption{Points $(u_1,u_2) \in \mathcal{D}_2$ may be mapped to the interior of the ellipse $u_1^2+u_1 u_2+u_2^2=1$ by defining $(v_1,v_2)=\frac{1}{u_1^2+u_2^2+u_1 u_2}(u_1,u_2)$. In this figure, we present the initial epochs of a number of consecutive eras in the five-dimensional BKL dynamics triggered by electric walls, in terms of the parametrization $(v_1,v_2)$. As it turns out, all such points $(v_1,v_2)$ lie within the region $\mathcal{I}$ delimited by $v_1^2+v_1 v_2+v_2^2 \leq 1$, $v_2\geq v_1$ and $v_1^2 + v_2^2 + v_1 v_2 + v_2 + 2  v_1 \leq 0$. We have plotted around 11k eras in the left figure and around 45k eras in the right plot. We observe how the interior of the ellipse gets uniformly filled, which conforms a hallmark of chaotic behavior. Observe that the density of initial points of eras change drastically around the dotted blue line $v_2=v_1+1$. Interestingly enough, we have checked that this is due to the fact that the initial epochs of eras whose previous era featured Kasner season II (see Definition \ref{def:ks}) are only allowed to be below the line $v_2=v_1+1$. Amusingly, numerically one may compute that average number of epochs in a given Kasner era is around $\sim$ 4.4.}
\label{fig:kasnerseasons5chaos}
\end{figure}

Specifically, let us show the direct relation between the free motion in the parameter space $\left\lbrace \dfrac{f_j}{\rho} \right\rbrace_{j=0}^{D-3}$ and in a hyperbolic billiard. Just like in the $D=4$ case, we define the logarithmic scale factors $\left\lbrace \beta_{i}=-\frac{1}{2} \log g_{ii} \right \rbrace_{i=0}^{D-2}$:
\begin{equation}
    \beta_i=\frac{\rho}{2}+\frac{(D-2-i) a_i f_i}{2}- \frac{1}{2} \sum_{j=0}^{i-1} a_j f_j\,, \quad i=0, \dots, D-2\,.
\end{equation}
The dominant walls creating the hyperbolic billiard in the hyperbolic space defined by projecting out the radial part of the $\left\lbrace \beta_{i}\right \rbrace_{i=0}^{D-2}$ coordinates arise from the $(D-1)$ gauge fields. By direct inspection to the electric potentials  \eqref{eq:potsol}, and comparing to the seminal work \cite{Damour:2002et}, one may see that motion will be constrained to the interior of the following region of the space of $\left\lbrace \beta_{i}\right \rbrace_{i=0}^{D-2}$:
\begin{equation}
    \beta_i \geq 0\,, \quad i=0,\dots, D-2\,.
\end{equation}
This region coincides exactly with the hyperbolic billiard that is formed by the electric walls considered in \cite{Damour:2002et}. For any space-time dimension $D$, these electric walls give rise to a compact hyperbolic billiard and, consequently, will give rise to chaotic dynamics --- see Figure \ref{fig:kasnerseasons5chaos} for an illustrative check of chaos in $D=5$. This is to be compared to the case of gravitational walls in higher-dimensional homogeneous cosmological models, for which chaos ceases to appear for $D>10$.

\begin{figure}
    \includegraphics[height=0.38\linewidth]{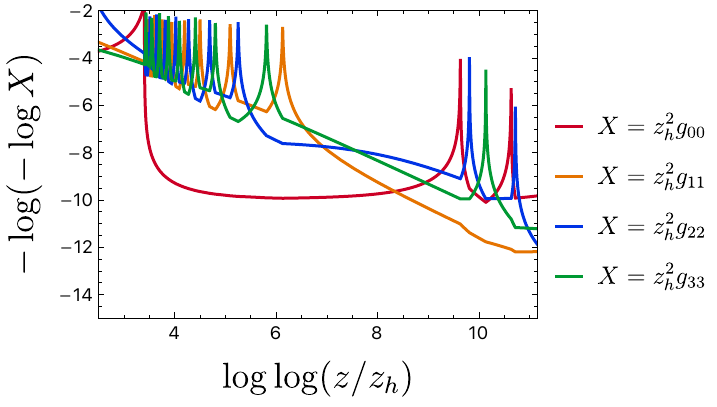}
     \includegraphics[width=0.45\linewidth]{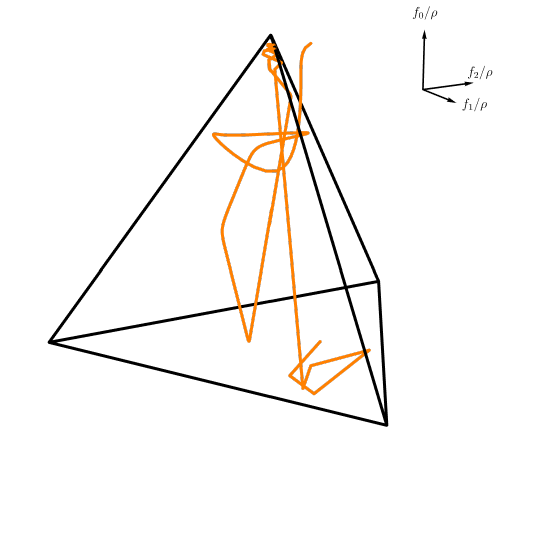}
    \caption{Mixmaster dynamics inside a five-dimensional AdS black hole. \textbf{Left}: The plot shows how the four independent metric components evolve with interior time. At late times, the evolution breaks up into the familiar BKL sequence of Kasner epochs, organized into eras. This is the direct generalization of the plots that may be found in the four-dimensional case \cite{DeClerck:2023fax}. \textbf{Right}: We explicitly observe that the dynamical evolution towards the singularity may be conceived as a set of free trajectories in the configuration space $\left\lbrace \frac{f_0}{\rho},\frac{f_1}{\rho},\frac{f_2}{\rho} \right\rbrace$ confined to the interior of a tetrahedron.}
    \label{Mixmaster5d}
\end{figure}


\subsubsection{Bouncing rules for Kasner exponents}

Let us now derive the collision or bouncing rules that specify the relation between the Kasner exponents of a given Kasner epoch with those of the previous Kasner epoch, brutally forced to an end once the free trajectory hits a face of the $(D-2)$-dimensional simplex. For this, let us take a look to the full equations of motion \eqref{eq:eom1} and \eqref{eq:eom2}. Direct computation reveals that:
\begin{equation}
\label{eq:rebotes}
    \frac{\ddot{f}_k}{\ddot{f}_0}=\frac{\frac{1}{V} \frac{\partial V}{\partial f_k}-\dot{f}_k}{\frac{1}{V} \frac{\partial V}{\partial f_0}-\dot{f}_0}\,, \quad k=1,\dots,D-3\,.
\end{equation}
Assume the bounce occurs against the wall ${\rm W}_m$, cf. \eqref{eq:dwalls}. In the infinite-well approximation, we may assume that all $\left\lbrace f_j \right\rbrace_{j=0}^{D-3}$ are continuous during the bounce, while their first derivatives suffer a jump. As a consequence, the potential $V$ and derivatives thereof may be taken to be constant during the bounce. In particular, the quantities:
\begin{equation}
    \lambda_j^{(m)}=\left. \frac{1}{V} \frac{\partial V}{\partial f_j} \right \vert_{{\rm W}_{m}} = \begin{cases}
a_j \,, \quad j< m\,,\\
-(D-2-j) a_j\,, \quad j=m\,, \\
0\,, \quad j>m\,,
\end{cases}
\end{equation}
will also be constant during the bounce. Consequently, if we consider equations \eqref{eq:rebotes} in the vicinity of a bounce against the $m$-th wall ${\rm W}_m$, one may integrate them to get 
\begin{equation}\label{bouncerule}
    \frac{\dot{f}_k-\lambda_k^{(m)}}{\dot{f}_k^{(0)}-\lambda_k^{(m)}}=    \frac{\dot{f}_0-\lambda_0^{(m)}}{\dot{f}_0^{(0)}-\lambda_0^{(m)}}\,, \quad k=1\dots, D-3\,, 
\end{equation}
where $\{\dot{f}_j^{(0)}\}_{j=0}^{D-3}$ denote the values of $\{\dot{f}_j\}_{j=0}^{D-3}$ immediately before the bounce. 

Observe that \eqref{bouncerule} conform a system of $D-3$ independent equations. Together with the quadratic constraint \eqref{eq:deff}, this defines a system of $D-2$ equations that determine the values of the $(D-2)$ parameters $\{\dot{f}_j\}_{j=0}^{D-3}$ after the bounces, assuming the knowledge of $\{\dot{f}_j^{(0)}\}_{j=0}^{D-3}$. The system of equations is conformed by $D-3$ linear equations and a  quadratic one, so the system will only have two unique sets of solutions: the trivial solution $\dot{f}_j=\dot{f}_j^{(0)}$ for $j=0,\dots, D-3$ (no bounce at all), and another one which represents the physical solution we are looking for (the values of the parameters after an actual bounce).

The physical solution may be expressed in a strikingly simple form in terms of the Kasner exponents $p_i$, which are related to the parameters $\dot{f}_j$ as dictated in \eqref{eq:kasnerxp}. After some cumbersome algebraic computations, one may show that such a physical solution against the $m$-th wall, with $(D-2) \geq m \geq 0$, can be obtained through the extremely simple system of linear equations:
\begin{equation}
  \frac{p_i}{p_m}=-a- \frac{p_i^{(0)}}{p_m^{(0)}}\,, \quad a=\frac{2}{(D-3)}\,, \quad i,m=0,\dots, D-2\,,\quad i \neq m\,,
  \label{eq:kasressis}
\end{equation}
where $\left \lbrace p_i^{(0)}, p_i \right\rbrace_{i=0}^{D-2}$ stand for the Kasner exponents before and after the bounce --- so that  both sets of exponents must satisfy the constraints \eqref{eq:kasnercons}. In $D=4$, these rules can be seen to match exactly with those in \cite{DeClerck:2023fax} after some algebraic work\footnote{Indeed, note their four-dimensional rules do not involve the Kasner exponent associated with the wall on which the bounce occurs, so one needs to massage their expressions to arrive to \eqref{eq:kasressis}.}. From here, one may solve \eqref{eq:kasressis} exactly to obtain:
\begin{equation}
\label{eq:kassol}
p_i=\frac{p_i^{(0)}+a \,p_m^{(0)}}{1+a \, p_m^{(0)}}\,, \quad p_m=-\frac{p_m^{(0)}}{1+a \, p_m^{(0)}}\,, \quad a=\frac{2}{D-3}\,.
\end{equation}
Interestingly enough, these bouncing rules for the Kasner exponents coincide with the ones that were found in \cite{Benini:2005su} in the context of homogeneous cosmological models with vector fields. 


Kasner exponents may be parametrized in terms of $(D-3)$ unconstrained parameters $\left\lbrace u_j\right\rbrace_{j=1}^{D-3}$ in a quite simple way. To this aim, take the following domain of definition $\mathcal{D}_{D-3}$ for these parameters:
\begin{equation}
\label{eq:dregion}
    \mathcal{D}_{D-3}=\left\lbrace (u_1,\dots,u_{D-3}) \in \mathbb{R}^{D-2} \left \vert\,  \lambda \geq 2+ \gamma \,, 1+\gamma \geq -u_{1}\,,  u_{D-3} \geq u_{D-2} \geq  \dots \geq u_1\right. \right \rbrace
\end{equation}
where we have defined:
\begin{equation}
 \lambda=\frac{1}{2} \left[1+\left (1+\sum_{j=1}^{D-3}u_j \right)^2+\sum_{j=1}^{D-3} u_j^2 \right]\,, \quad \gamma=\sum_{j=1}^{D-3}u_j\,.
 \label{eq:deflg}
\end{equation}
Define the following :
\begin{equation}
\label{eq:parkas}
    \mathfrak{p}_{D-2}=\frac{\lambda-1}{\lambda}\,, \quad \mathfrak{p}_{D-3}=\frac{1+ \gamma}{\lambda} \,, \quad \mathfrak{p}_{D-3-j}=\frac{-u_j}{\lambda}\,, \quad j=1,\dots,D-3\,.
\end{equation}
By construction, $\mathfrak{p}_{D-2} \geq \mathfrak{p}_{D-3} \geq \dots \geq \mathfrak{p}_1 \geq \mathfrak{p}_0$ and naturally satisfy $\sum_{i=0}^{D-2}\mathfrak{p}_i^2=\sum_{i=0}^{D-2}\mathfrak{p}_i=1$. Therefore, given any set of Kasner exponents $\left\lbrace p_i \right\rbrace_{i=0}^{D-2}$ satisfying that $p_{D-2} \geq p_{D-3} \geq \dots p_{1} \geq p_{0}$ (maybe after a permutation),  we may set $p_i=\mathfrak{p}_{i}$ for every $i=0,\dots,D-2$.
This parametrization is a equivalent\footnote{If $\left\lbrace u_j^{\rm EH}\right\rbrace_{j=1}^{D-3}$ stands for the parameters in \cite{Elskens:1987rk,Elskens:1987gj}, then  $u_j^{\rm EH}=-u_{D-3-j}$.} to that presented in \cite{Elskens:1987rk,Elskens:1987gj} and generalizes the original four-dimensional parametrization of Kasner exponents provided in \cite{Belinsky:1970ew}. 


\subsection{Kasner dynamics in $D=5$: introducing Kasner seasons}

In order to gain a better understanding of Kasner dynamics in higher-dimensional black holes, we will first consider the case $D=5$, which will already showcase intriguing distinctive features with respect to the four-dimensional case.

In $D=5$, Kasner exponents are controlled by two parameters $(u_1,u_2)$ defined in the two-dimensional region --- just set $D=5$ in \eqref{eq:dregion}: 
\begin{equation}
\label{eq:5region}
\mathcal{D}_2=\left\lbrace (u_1,u_2) \in \mathbb{R}^2 \, \vert \,  u_1^2+u_1 u_2 +u_2^2 \geq 1 \,, u_2 \geq u_1\,, u_2 \geq -2u_1-1\right\rbrace\,.
\end{equation}
The parametrization \eqref{eq:parkas} in $D=5$ reads as follows:
\begin{align}
\nonumber
    \mathfrak{p}_{3}=1-\frac{1}{\lambda(u_1,u_2)}\,, \quad &  \mathfrak{p}_{2}=\frac{1+u_1+u_2}{\lambda(u_1,u_2)} \quad  \mathfrak{p}_1=-\frac{u_1}{\lambda(u_1,u_2)}\,, \quad  \mathfrak{p}_{0}=-\frac{u_2}{\lambda(u_1,u_2)}\,,  \\ \lambda(u_1,u_2)&=1+u_1+u_1^2+u_1 u_2+u_2+u_2^2\,,
    \label{eq:parkas5}
\end{align}
Within the region $\mathcal{D}_2$ defined in \eqref{eq:5region}, note that $\mathfrak{p}_3 \geq \mathfrak{p}_2 \geq \mathfrak{p}_1 \geq \mathfrak{p}_0$. Also, using \eqref{eq:rangokas}, we have that:
\begin{equation}
    1 \geq \mathfrak{p}_{3} \geq \frac{1}{2}\,, \quad \frac{1+\sqrt{3}}{4} \geq \mathfrak{p}_{2} \geq 0\,, \quad \frac{1}{2} \geq \mathfrak{p}_{1} \geq \frac{1-\sqrt{3}}{4}\,, \quad 0 \geq \mathfrak{p}_{0} \geq -\frac{1}{2}\,.
    \label{eq:rangoexp5}
\end{equation}
As a consequence, we see that there will always be, at least, two non-negative Kasner exponents (therefore, at most two negative Kasner exponents) and a maximum\footnote{Of course, there is the trivial case of four non-negative Kasner exponents given by $\mathfrak{p}_{3}=1$ and $\mathfrak{p}_2=\mathfrak{p}_1=\mathfrak{p}_{0}=0$.} of three positive Kasner exponents (respectively, a minimum of a single negative Kasner exponent).

Assume a bounce occurs in the wall associated with the exponent $p_m$ (so that $p_m$ is the least Kasner exponent). Let $p_J=\left\lbrace p_j,p_k,p_l,p_m\right\rbrace$ denote the Kasner exponents before the bounce. Let $p_J'=\left\lbrace p_j',p_k',p_l',p_m'\right\rbrace$ stand for the exponents after the bounce. By setting $D=5$ in \eqref{eq:kassol} we get the following collision law $\mathcal{E}_5$:
\begin{equation}
   \label{eq:kassol5} 
   \mathcal{E}_5: \left (p_j,p_k,p_l,p_m \right) \rightarrow \left (\frac{p_j+p_m}{1+p_m},\frac{p_k+p_m}{1+p_m},\frac{p_l+p_m}{1+p_m},-\frac{p_m}{1+p_m} \right)
\end{equation}
Let us examine in close detail the new Kasner exponents after the bounce. Assume that $p_j>p_k>p_l>p_m$ (the case in which two Kasner exponents are equal will be succinctly studied in Section \ref{subsec:restricted_triangle}). As such, there exist parameters $(u_1,u_2) \in \mathcal{D}_2$ such that $p_j=\mathfrak{p}_{3}(u_1,u_2)$, $p_k=\mathfrak{p}_{2}(u_1,u_2)$, $p_l=\mathfrak{p}_{1}(u_1,u_2)$ and $p_m=\mathfrak{p}_{0}(u_1,u_2)$. Since $p_m<0$, we necessarily have that $p_m'>0$. Depending on the initial values for the exponents, the new Kasner exponents \eqref{eq:kassol5} will be determined by some new parameters $(u_1',u_2') \in \mathcal{D}_2$ and a new reordering $p'_{\sigma_A(j)}>p'_{\sigma_A(k)}>p'_{\sigma_A(l)}>p'_{\sigma_A(m)}$ for a given permutation $\sigma_A$. As a consequence, the collision law $\mathcal{E}_5$ given by \eqref{eq:kassol5} may be decomposed into a \emph{BKL law} $\mathcal{K}_A$ and a \emph{BKL reordering} $\sigma_A$ 
\begin{equation}
    \mathcal{E}_5: \{p_j,p_k,p_l,p_m\} \rightarrow \left\lbrace p_j',p_k',p_l',p_m' \right\rbrace\longleftrightarrow \left\lbrace \begin{matrix}
        \mathcal{K}_A: \mathcal{D}_2 \rightarrow \mathcal{D}_2\\
        \sigma_A: \mathbb{Z}_4 \rightarrow \mathbb{Z}_4
    \end{matrix} \right.\,, \quad A={I,II,III}
\end{equation}
according to three different cases, as described in Table \ref{tabla:kasnerexp}. For definiteness, the composite function $\mathcal{B}_A: (\mathcal{K}_A, \sigma_A): \mathcal{D}_2 \times \mathbb{Z}_4 \rightarrow  \mathcal{D}_2 \times \mathbb{Z}_4$ will be called the \emph{BKL map}.

\begin{table}[t!]
\centering
\renewcommand{\arraystretch}{2}

\begin{tabular}{|c|c|c|c|}
\cline{1-4}
\textbf{BKL map} $\mathcal{B}_A$ & $\mathcal{B}_I$ &  $\mathcal{B}_{II}$ & $\mathcal{B}_{III}$ \\ 
\hline
{\textbf{Range of}} & $u_2>1+u_1\,$ and & $u_2<1+u_1\,$ and & \multirow{2}{*}{$\mathcal{C}(u_1,u_2)< 1$}\\
{\textbf{Application}} & $\mathcal{C}(u_1,u_2)> 1$ & $\mathcal{C}(u_1,u_2)> 1$ &  \\
\hline
\textbf{BKL law} & $u_1'=-1-u_1$ & $u_1'=-u_2$ & $u_1'=\dfrac{-1-u_1}{\mathcal{C}(u_1,u_2)}$ \\ $\mathcal{K}_A(u_1,u_2)=(u_1',u_2')$ & $u_2'=u_1+u_2$ & $u_2'=u_1+u_2$ & $u_2'=\dfrac{u_1+u_2}{\mathcal{C}(u_1,u_2)}$ \\
\hline
\textbf{BKL reordering}  & \multirow{2}{*}{$\sigma_I(J)=\{j,m,k,l\}$}& \multirow{2}{*}{$\sigma_{II}(J)=\{j,k,m,l\}$}& \multirow{2}{*}{$\sigma_{III}(J)=\{m,j,k,l\}$} \\ $p_{\sigma(J)}'=\mathfrak{p}_{N}(u_1',u_2')$ & & &  \\
\hline
\end{tabular}
\caption{New Kasner exponents $
p'_J$ in terms of the initial ones $
p_J$, where $J=\{j,k,l,m\}$ and $N=\{3,2,1,0\}$. The permutation $\sigma_A(J)$ indicates the new ordering of Kasner exponents (from larger to smaller). It is assumed that $p_j >p_k>p_l>p_m$. We defined $\mathcal{C}(u_1,u_2)=1+u_1^2+u_2^2+u_1 u_2+u_1-u_2$. Note that if $(u_1,u_2) \in \mathcal{D}_2$, then $(u_1',u_2') \in \mathcal{D}_2$ for all three Kasner maps.  We do not consider the case of strict equalities in the row \emph{Range of Application} as this corresponds to having two equal Kasner exponents --- which we are not considering in this section.   }
\label{tabla:kasnerexp}
\end{table}

As before, periods of definite and fixed Kasner exponents are called \emph{Kasner epochs}, that come abruptly to an end and transition to a new Kasner epoch by application of any of the Kasner maps $\mathcal{B}_A$ in Table \ref{tabla:kasnerexp}. These Kasner epochs may be encompassed in \emph{Kasner eras}, defined as those set of Kasner epochs in which the direction of the largest Kasner exponent is held constant. By direct inspection to Table \ref{tabla:kasnerexp}, we observe that the BKL map $\mathcal{B}_{III}$ corresponds naturally to a change of Kasner era, as it is the unique transformation in which the largest Kasner direction becomes swapped after the bounce. On the other hand, the BKL maps $\mathcal{B}_I$ and $\mathcal{B}_{II}$ correspond to transitions between Kasner epochs within the same era. These produce inequivalent reorderings while maintaining the direction of the largest Kasner exponent. This in sharp contrast to the four-dimensional case, where we only had a single BKL map within a given era. 

These novel higher-dimensional features inspire us to define the notion of a \emph{Kasner season}\footnote{One may be worried about the fact that \emph{season} does not start with the letter \emph{e}, as it is the case with epoch, era and eon. We would like to note that in Spanish this would be indeed the case (Kasner season translates as \emph{estaci\'on Kasner}, thus respecting the \emph{e}-rule).}.
\begin{defi*}
  Given a certain Kasner epoch, we will say that it is in the Kasner season $A$, where $A$ belongs to some index set, if the following epoch is obtained by application of the BKL map $\mathcal{B}_A$.
  \label{def:ks}
\end{defi*}

In our five-dimensional case, we may identify three different Kasner seasons associated to the three BKL maps $\mathcal{B}_A$, with $A=\{I,II,III\}$. On the one hand, if a Kasner epoch experiences Kasner season III, it implies that it is the last epoch of a given era. On the other hand, Kasner seasons I and II corresponds to transitions within the same era. Note that only two different Kasner seasons appear in the standard four-dimensional BKL setup: one season associated to transitions of epochs within an era, and another season corresponding to change of eras. Since the second Kasner season in four dimensions only corresponds to change or eras, the dynamics of Kasner seasons is equivalent to the evolution of Kasner eras. However, in higher dimensions, we have more than one inequivalent  BKL map driving transitions of epochs within the same era, so that dynamics of Kasner seasons is manifestly different from that of eras.  




Let us consider Kasner eras conformed by more than one epoch. Let $(u_1,u_2)$ denote the parameters associated with some initial set of Kasner exponents $\{p_i\}_{i=1}^4$ at the beginning of a Kasner era, which  will satisfy $\mathcal{C}\left (u_1,u_2\right)=\lambda\left (u_1,u_2 \right)-2u_2 >1$. Also, we will denote by $\left (u_1^{(n)},u_2^{(n)} \right )$ the parameters characterizing the $(n+1)$-th epoch of the era. We may distinguish the three following qualitative behaviors within such Kasner eras:
\begin{enumerate}
    \item[\textbf{Case 1:}] Assume that $\dfrac{1}{\sqrt{3}}> u_1 > -1-\dfrac{1}{\sqrt{3}}$. Then, the Kasner era will consist of the consecutive application of BKL map $\mathcal{B}_I$ $n \geq 1$ times (cf. Table \ref{tabla:kasnerexp}) until $\mathcal{C}\left (u_1^{(n)},u_2^{(n)}\right)<1$. This Kasner epoch will conclude with the application of BKL map $\mathcal{B}_{III}$, and the start of a new era. Consequently, the era will feature $n$ epochs in Kasner season I followed by one epoch in Kasner season III.
    \item[\textbf{Case 2:}] Now, assume that $u_1>\dfrac{1}{\sqrt{3}}$ or that $ u_1 < -1-\dfrac{1}{\sqrt{3}}$. Furthermore, suppose that $u_2>1+u_1$. In such a case, the Kasner era will start with the iterative application of BKL map $\mathcal{B}_I$ in Table \ref{tabla:kasnerexp} $n \geq 1$ times, until the corresponding parameters $\left (u_1^{(n)},u_2^{(n)} \right)$ are such that $1+u_1^{(n)}>u_2^{(n)}>u_1^{(n)}$. At this point, we will still have $\mathcal{C}\left (u_1^{(n)},u_2^{(n)} \right)>1$ and one needs to apply now BKL map $\mathcal{B}_{II}$,  so that $u_1^{(n+1)}=-u_2^{(n)}$ and $u_2^{(n+1)}=u_1^{(n)}+u_2^{(n)}$. Since $u_2^{(n+1)}>1+u_1^{(n+1)}$, if $\mathcal{C}\left (u_1^{(n+1)},u_2^{(n+1)} \right)>0$, BKL map $\mathcal{B}_I$ needs to be used --- otherwise, BKL map $\mathcal{B}_{III}$ would apply and the era comes to an end. One can get convinced that $\mathcal{C}\left (u_1^{(n+2)},u_2^{(n+2)} \right)>1$ and $1+u_1^{(n+2)}>u_2^{(n+2)}>u_1^{(n+2)}$, so BKL map $\mathcal{B}_{II}$ is of use again. This process goes on until after some application of BKL map $\mathcal{B}_{II}$ provides $\mathcal{C}\left (u_1^{(n+m)},u_2^{(n+m)} \right)<1$ for certain even integer $m\geq 0$. Then one needs to apply BKL map $\mathcal{B}_{III}$, leading to a new Kasner era. Therefore, this type of eras initiates with $n$ epochs belonging to Kasner season I, consecutive epochs of alternating seasons II and I the final epoch in the Kasner season III.

    \item[\textbf{Case 3:}] If $u_1>\dfrac{1}{\sqrt{3}}$ or $ u_1 < -1-\dfrac{1}{\sqrt{3}}$, but $u_2<1+u_1$, then the Kasner era will begin with the application of BKL map $\mathcal{B}_{II}$. If $\mathcal{C}\left (u_1^{(1)},u_2^{(1)} \right)>1$, then one needs to apply iteratively BKL maps  $\mathcal{B}_{I}$ and $\mathcal{B}_{II}$ until $\mathcal{C}\left (u_1^{(1+m)},u_2^{(1+m)} \right)<1$ for an even integer $m \geq 0$. Direct application of BKL map $\mathcal{B}_{III}$ concludes the Kasner era. These eras consist of successive epochs of Kasner seasons II and I followed by another epoch in season II and a last epoch in the season III.
    
    
\end{enumerate}

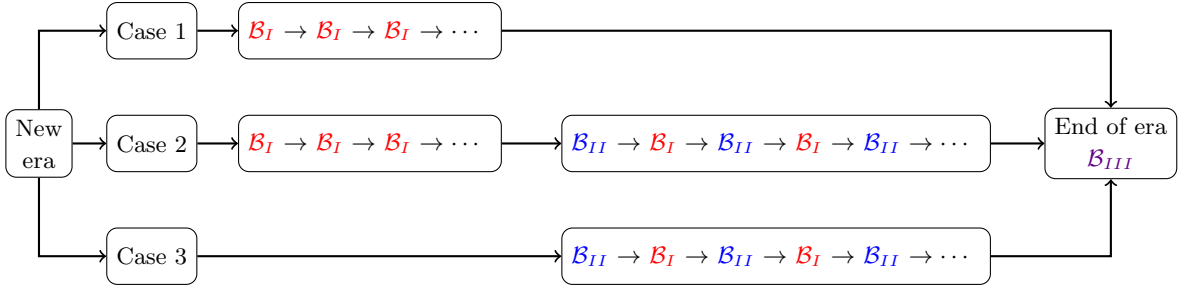
\begin{figure}
    \centering
\begin{tikzpicture}[
   node distance=1.8cm and 2.5cm, font=\footnotesize,
    box/.style={
        draw, rounded corners,
        minimum width=0.75cm,
        minimum height=0.75cm,
        align=center
    }
]

\node[box] (newera)  {New\\era};

\node[box, xshift=1.5cm, yshift=1.5cm] (case1) {Case 1};
\node[box, xshift=1.5cm] (case2) {Case 2};
\node[box, xshift=1.5cm, yshift=-1.5cm] (case3) {Case 3};

\node[box, xshift=4.4cm, yshift=1.5cm] (k11) {
\begingroup \footnotesize
{\color{red} $\mathcal{B}_{I}$} $\rightarrow$ {\color{red}$\mathcal{B}_{I}$} $\rightarrow$ {\color{red}$\mathcal{B}_{I}$}  $\rightarrow \cdots$ \endgroup};
\node[box, xshift=4.4cm] (k21) {
\begingroup \footnotesize
{\color{red}$\mathcal{B}_{I}$} $\rightarrow$ {\color{red}$\mathcal{B}_{I}$} $\rightarrow$ {\color{red}$\mathcal{B}_{I}$}  $\rightarrow \cdots$ \endgroup};

\node[box, xshift=9.8cm] (k22) {\begingroup \footnotesize
{\color{blue}$\mathcal{B}_{II}$} $\rightarrow$ {\color{red}$\mathcal{B}_{I}$} $\rightarrow$ {\color{blue}$\mathcal{B}_{II}$} $\rightarrow$ {\color{red}$\mathcal{B}_{I}$} $\rightarrow$ {\color{blue}$\mathcal{B}_{II}$} $\rightarrow \cdots$
\endgroup};
\node[box, xshift=9.8cm, yshift=-1.5cm] (k32) {
\begingroup \footnotesize
{\color{blue}$\mathcal{B}_{II}$} $\rightarrow$ {\color{red}$\mathcal{B}_{I}$} $\rightarrow$ {\color{blue}$\mathcal{B}_{II}$} $\rightarrow$ {\color{red}$\mathcal{B}_{I}$} $\rightarrow$ {\color{blue}$\mathcal{B}_{II}$} $\rightarrow \cdots$
\endgroup};

\node[box,  xshift=14.25cm] (end) {End of era \\ \begingroup  \footnotesize
{\color{purple!80!black!95!red!60!blue}{$\mathcal{B}_{III}$}}
\endgroup
};

\draw[->, thick]
    (newera.north) -- (0,1.5cm) -- (case1);
\draw[->, thick] (newera) -- (case2);
\draw[->, thick]
    (newera.south) -- (0,-1.5cm) -- (case3);

\draw[->, thick] (case1) -- (k11);
\draw[->, thick] (case2) -- (k21);

\draw[->, thick] (k21) -- (k22);
\draw[->, thick] (case3) -- (k32);

\draw[->, thick] (k11) -- (14.25,1.5cm) -- (end.north);
\draw[->, thick] (k22) -- (end);
\draw[->, thick] (k32) -- (14.25,-1.5cm) -- (end.south);

\end{tikzpicture}
\caption{Structure of a Kasner era made of more than one epoch, depending on the initial values of the parameters $(u_1,u_2)$ at the beginning of the era --- divided into the cases 1, 2 and 3 as described in the main text. In Case 1, the era only features the Kasner season I until Kasner season III, which implies the end of the era. In Case 2, the era begins with iterative epochs in Kasner season I. At some point, the following epochs will belong to Kasner seasons II and I alternatively, until an epoch in Kasner season III arises and the era concludes. Finally, in Case 3 the era displays a continuous exchange between Kasner seasons II and I until the era comes to an end with an epoch in Kasner season III. We have colored the different BKL maps to remark the different patterns of eras.}
\label{fig:ksd}
\end{figure}

The various types of eras are represented schematically in Figure \ref{fig:ksd}.

\subsubsection{Kasner dynamics in $D=5$: the case of gravitational walls}

So far, we have analyzed how Kasner exponents get modified after bouncing against electric walls. However, in homogeneous models without $p$-forms, the dominant walls in the approach towards the singularity will correspond to gravitational walls. In $D=4$, the bouncing rules associated to electric or gravitational walls are exactly the same. This will no longer be the case for $D \geq 5$, motivating us to revisit the corresponding bouncing rules, explain the appearance of Kasner seasons and find an intriguing relation between the bouncing rules arising from electric or gravitational walls.

To this aim, let us delve into the study of Kasner dynamics associated with gravitational walls in $D=5$. Take $q_J=\{q_j,q_k,q_l,q_m\}$ to be the Kasner exponents before the bounce and let $q_J'=\{q_j',q_k',q_l',q_m'\}$ be the exponents afterwards. We will consider the generic case in which $q_j > q_k >q_l> q_m$. The collision law $\mathcal{G}_5$ that relates $q_J$ and $q_J'$ is given by \cite{Demaret:1986ys}:
\begin{equation}
   \mathcal{G}_5: \left( q_j,q_k,q_l,q_m   \right) \rightarrow \left ( \frac{1-q_k+q_m}{\Delta_q}, \frac{1-q_j+q_m}{\Delta_q}, \frac{q_l}{\Delta_q}, \frac{q_j+q_k-1}{\Delta_q}\right) \,, 
   \label{eq:kasexpgrav}
\end{equation}
where we defined $\Delta_q=2-q_j-q_k+q_m$. It can be easily proven that $q'_j > q'_m> q'_k$. Just like in the case with electric walls, let us consider parameters $(u_1,u_2) \in \mathcal{D}_2$ such that $q_j=\mathfrak{p}_3(u_1,u_2)$, $q_k=\mathfrak{p}_2(u_1,u_2)$, $q_l=\mathfrak{p}_1(u_1,u_2)$ and $q_m=\mathfrak{p}_0(u_1,u_2)$. As it turns out, the collision law $\mathcal{G}_5$ \eqref{eq:kasexpgrav} can be translated into four gravitational BKL maps $\mathcal{B}_C^{(g)}$, with $C=\{I,II,III,e_1,e_2,e_3,e_4,e_5\}$, comprising BKL laws $\mathcal{K}_C^{(g)}$ and BKL reorderings $\sigma_C^{(g)}$ as follows:
\begin{align}
    \mathcal{G}_5: &\{q_j,q_k,q_l,q_m\} \rightarrow \left\lbrace q_j',q_k',q_l',q_m' \right\rbrace\longleftrightarrow \left\lbrace \begin{matrix}
        \mathcal{K}_C^{(g)}: \mathcal{D}_2 \rightarrow \mathcal{D}_2\\
        \sigma_C^{(g)}: \mathbb{Z}_4 \rightarrow \mathbb{Z}_4
    \end{matrix} \right.\,, \\ \label{eq:Cindice} C&=\{I,II,III,e_1,e_2,e_3,e_4,e_5\}
\end{align}
The actions of the various gravitational BKL maps are summarized in Table \ref{tabla:kasnerexpg} and \ref{tabla:kasnerexpgera}. With gravitational walls, we observe that Kasner epochs may be in eight different (gravitational) Kasner seasons $C$ (with $C$ as in \eqref{eq:Cindice}), depending on which BKL map needs to be applied afterwards. As it is evident from Table \ref{tabla:kasnerexpg}, Kasner seasons I, II and III may take place for those Kasner epochs that do not lead into a new era, while Kasner epochs $\vec{e}=\{e_1,e_2,e_3,e_4,e_5\}$ correspond to the various seasons associated to the different possible changes of eras and the subsequent new reorderings of exponents.

Kasner eras conformed by only one epoch are rather trivial: they just transition towards the following era according to the the Kasner maps $\mathcal{B}_{\vec{e}}^{(g)}$ --- cf. Table \ref{tabla:kasnerexpgera}. Therefore, let us concentrate on those eras containing more than one epoch. If $(u_1,u_2) \in \mathcal{D}_2$ represents the initial epoch of the era, which will satisfy by construction $\mathcal{C}\left (u_1,u_2\right)=\lambda\left (u_1,u_2 \right)-2u_2 >1$, and $\left(u_1^{(n)},u_2^{(n)}\right)$ denotes the $(n+1)$-th epoch of the era, we may distinguish the following cases:
\begin{enumerate}
    \item[\textbf{Case 1:}] Assume $\dfrac{1}{\sqrt{3}}> u_1 > -1-\dfrac{1}{\sqrt{3}}$. Then, the Kasner era will be conformed by consecutive applications of BKL map $\mathcal{B}_I^{(g)}$ $n \geq 1$ times (see Table \ref{tabla:kasnerexpg}) until $\mathcal{C}\left (u_1^{(n)},u_2^{(n)}\right)<1$, where $\left ( u_1^{(n)},u_2^{(n)} \right)$ stand for the parameters of the Kasner exponents associated to the $(n+1)$-th Kasner epoch in the era. Next, a BKL map $\mathcal{B}_{\vec{e}}^{(g)}$ applies and a new era begins --- the specific one is selected according to the criteria in Table \ref{tabla:kasnerexpgera}. The era will feature $n$ epochs in Kasner season I followed by one epoch in one of the Kasner seasons $\vec{e}$.
    \item[\textbf{Case 2:}] Now, assume that $u_1>\dfrac{1}{\sqrt{3}}$ and
    $u_2>1+u_1$. The Kasner era will proceed with the iterative application of BKL map $\mathcal{B}_I^{(g)}$ (cf. Table \ref{tabla:kasnerexpg}) $n \geq 1$ times, arriving to a halt when the corresponding parameters $\left (u_1^{(n)},u_2^{(n)} \right)$ are such that $1+u_1^{(n)}>u_2^{(n)}>u_1^{(n)}$. At this point, it still holds that $\mathcal{C}\left (u_1^{(n)},u_2^{(n)} \right)>1$ and the era continues by the application of the BKL map $\mathcal{B}_{II}^{(g)}$ $m \geq 1$ times, until one gets $\mathcal{C}\left (u_1^{(n+m)},u_2^{(n+m)} \right)<1$. At this point, one must resort to Table \ref{tabla:kasnerexpgera} and use one of the BKL maps $\mathcal{B}_{\vec{e}}^{(g)}$, initiating a new Kasner era. As a result, in this case eras start with $n$ epochs belonging to Kasner season I, $m$ epochs in seasons II and a final epoch in one of the Kasner seasons $\vec{e}$.
    

    \item[\textbf{Case 3:}] Suppose that $u_1 < -1-\dfrac{1}{\sqrt{3}}$ and
    $u_2>-2u_1$. Now, the era has the same structure as in Case 2, exchanging $\mathcal{B}_{II}^{(g)}$ by $\mathcal{B}_{III}^{(g)}$ (\emph{mutatis mutandis}). 

    \item[\textbf{Case 4:}] If $u_1>\dfrac{1}{\sqrt{3}}$ but $u_2<1+u_1$, 
    the Kasner era will be made of consecutive applications of BKL map $\mathcal{B}_{II}^{(g)}$, until one gets  some parameters such that  $\mathcal{C}\left (u_1^{(m)},u_2^{(m)} \right)<1$ for an integer $m \geq 1$. Then, application of one of the BKL maps $\mathcal{B}_{\vec{e}}^{(g)}$, following the instructions in Table \ref{tabla:kasnerexpgera}, finishes the Kasner era. These eras consist of $m$ epochs in Kasner seasons II followed by a last epoch in one of the seasons $\vec{e}$. 

    \item[\textbf{Case 5:}] Assume $u_1 < -1-\dfrac{1}{\sqrt{3}}$ and
    $u_2<-2u_1$. This type of eras are completely analogous to eras of Case 4, changing $\mathcal{B}_{II}^{(g)}$ by $\mathcal{B}_{III}^{(g)}$. 
\end{enumerate}

\begin{table}[t!]
\centering
\renewcommand{\arraystretch}{2}

\begin{tabular}{|c|c|c|c|c|}
\cline{1-5}
\textbf{BKL map} $\mathcal{B}_C^{(g)}$& $\mathcal{B}_I^{(g)}$ & $\mathcal{B}_{II}^{(g)}$ & $\mathcal{B}_{III}^{(g)}$  & $\mathcal{B}_{\vec{e}}^{(g)}$ \\ 
\hline
{\textbf{Range of}} & $u_2 > -2u_1\,$  & $u_2> -2u_1\,$  & \multirow{2}{*}{$u_2< -2u_1$}  &\multirow{3}{*}{$\mathcal{C}(u_1,u_2)< 1$}  \\
& $u_2> 1+u_1$ & $u_2< 1+u_1$  & & \\
{\textbf{Application}} &$\mathcal{C}(u_1,u_2)> 1$ & $\mathcal{C}(u_1,u_2)> 1$ & $\mathcal{C}(u_1,u_2)> 1$ & \\
\hline
\textbf{BKL law} & $u_1'=u_1$ & $u_1'=u_2-1$ & $u_1'=-u_1-u_2$ & \multirow{2}{*}{See Table \ref{tabla:kasnerexpgera}}\\ $\mathcal{K}_C^{(g)}(u_1,u_2)=(u_1',u_2')$ & $u_2'=u_2-1$ & $u_2'=u_1$ & $u_2'=u_2-1$ &   \\
\hline
\textbf{BKL reordering}  & \multirow{2}{*}{$\{j,m,l,k\}$}& \multirow{2}{*}{$\{j,m,k,l\}$}& \multirow{2}{*}{$\{j,l,m,k\}$} & \multirow{2}{*}{See Table \ref{tabla:kasnerexpgera}} \\ $q_{\sigma^{(g)}(J)}'=\mathfrak{q}_{N}(u_1',u_2')$ & & & &  \\
\hline
\end{tabular}
\caption{New Kasner exponents $
q'_J$ in terms of the initial ones $
q_J$, where $J=\{j,k,l,m\}$ and $N=\{3,2,1,0\}$. It is assumed that $q_j >q_k>q_l>q_m$. As in Table \ref{tabla:kasnerexp},  $\mathcal{C}(u_1,u_2)=1+u_1^2+u_2^2+u_1 u_2+u_1-u_2$. Note that if $(u_1,u_2) \in \mathcal{D}_2$, then $(u_1',u_2') \in \mathcal{D}_2$ for all four BKL maps.} 
\label{tabla:kasnerexpg}
\end{table}

\begin{table}[t!]
\centering
\renewcommand{\arraystretch}{2}

\begin{tabular}{|c|c|c|c|c|c|}
\cline{1-6}
\textbf{BKL map} $\mathcal{B}_{\vec{e}}^{(g)}$& $\mathcal{B}_{e_1}^{(g)}$ & $\mathcal{B}_{e_2}^{(g)}$ & $\mathcal{B}_{e_3}^{(g)}$  & $\mathcal{B}_{e_4}^{(g)}$ &
$\mathcal{B}_{e_5}^{(g)}$
\\ 
\hline
{\textbf{Domain of}} & \multirow{2}{*}{$\mathcal{R}_1$}  & \multirow{2}{*}{$\mathcal{R}_2$}  & \multirow{2}{*}{$\mathcal{R}_3$}  & \multirow{2}{*}{$\mathcal{R}_4$} & \multirow{2}{*}{$\mathcal{R}_5$}  \\
{$(u_1,u_2)\in \mathcal{D}_2$} &  &  & & & \\
\hline
\textbf{BKL law} $\mathcal{K}_{\vec{e}}^{(g)}$ & $\mathcal{K}_{III} \circ \mathcal{K}_{II}$ & $\mathcal{K}_{III} \circ \mathcal{K}_I$ & $\mathcal{K}_{III} \circ \mathcal{K}_{III}$ & $\mathcal{K}_{I} \circ \mathcal{K}_{III}$ &
$\mathcal{K}_{II} \circ \mathcal{K}_{III}$\\
\hline
\textbf{BKL reordering}  &  \multirow{2}{*}{$\{m,j,k,l \}$} &  \multirow{2}{*}{$\{ m,j,l,k\}$} &  \multirow{2}{*}{$\{ m,l,j,k\}$} &  \multirow{2}{*}{$\{l,m,j,k \}$} &  \multirow{2}{*}{$\{l,j,m,k \}$} \\ $q_{\sigma(J)}'=\mathfrak{q}_{N}(u_1',u_2')$ & & & & &  \\
\hline
\end{tabular}
\caption{New Kasner exponents $
q'_J$ in terms of the initial ones $
q_J$, where $J=\{j,k,l,m\}$ and $N=\{3,2,1,0\}$, whenever the BKL maps $\mathcal{B}_{\vec{e}}^{(g)}$ apply. Indeed, it is observed that this correspond to various types of change of era. It is assumed that $q_j >q_k>q_l>q_m$. The regions $\{\mathcal{R}_s \}_{s=1}^5$ are defined in Figure \ref{fig:regiones_grav_est}.}
\label{tabla:kasnerexpgera}
\end{table}

\begin{figure}[t!]
\centering
\begin{minipage}{0.45\linewidth}
    \includegraphics[scale=0.28]{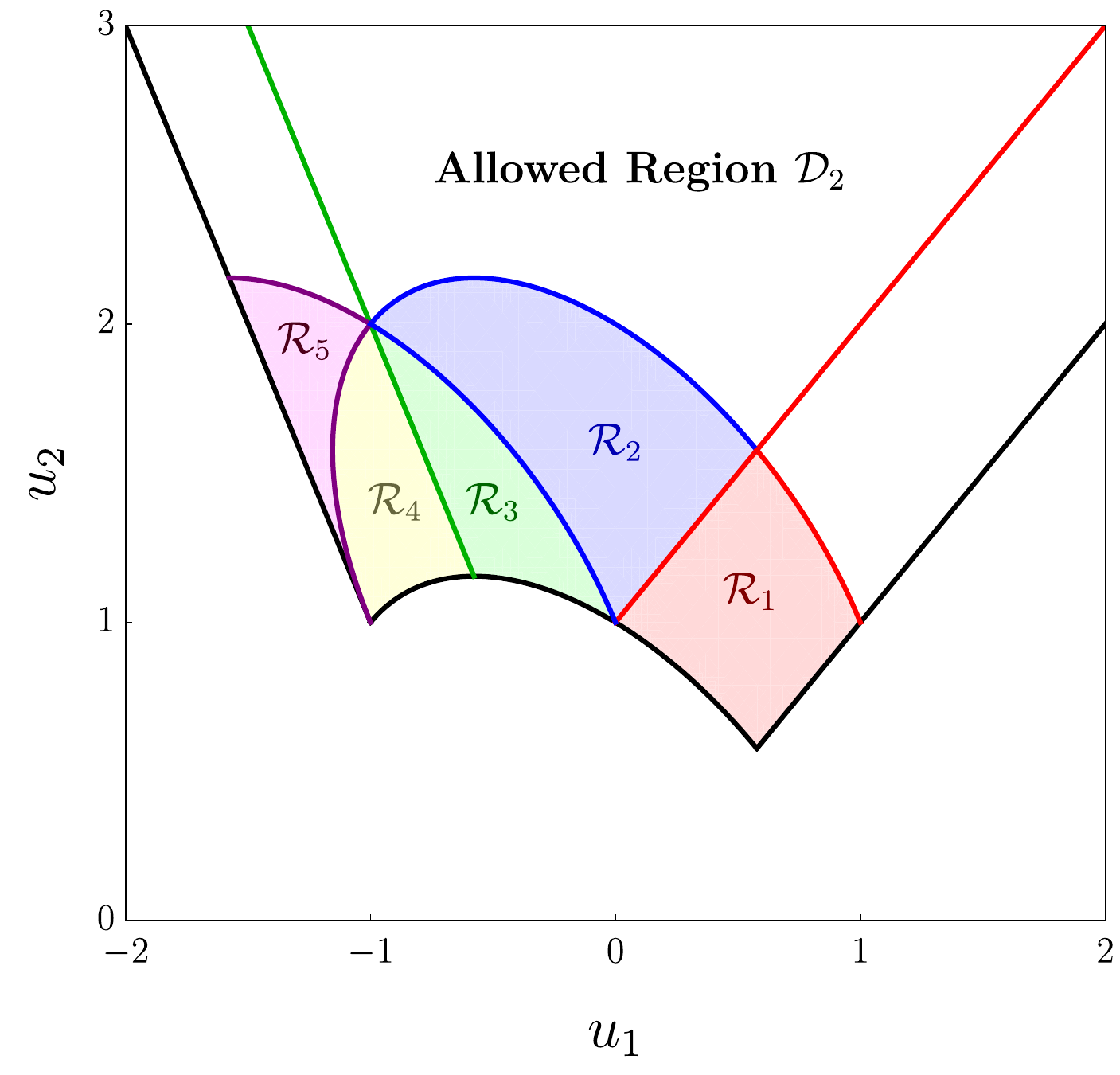}
\end{minipage}
\begin{minipage}{0.5\linewidth}
\small
    \begin{align*}
    \, \, \,\,  \mathcal{R}_1&=\left\lbrace (u_1,u_2) \in \mathcal{D}_2 \vert \, Q_1< 0\,, \, u_2\leq u_1+1 \right \rbrace\,,\\
 \, \, \,   \,  \mathcal{R}_2&=\left\lbrace (u_1,u_2) \in \mathcal{D}_2 \vert \, Q_1< 0\,, \, Q_2> 0\,, \, u_2\geq u_1+1 \right \rbrace\,, \\
   \, \, \, \,  \mathcal{R}_3&=\left\lbrace (u_1,u_2) \in \mathcal{D}_2 \vert \, Q_1< 0\,, \, Q_2< 0\,, \, u_2\geq -2u_1 \right \rbrace\,, \\
   \, \, \, \,  \mathcal{R}_4&=\left\lbrace (u_1,u_2) \in \mathcal{D}_2 \vert \, Q_1< 0\,, \, u_2\leq -2u_1 \right \rbrace\,, \\
  \, \, \,  \,  \mathcal{R}_5&=\left\lbrace (u_1,u_2) \in \mathcal{D}_2 \vert \, Q_1> 0\,, \, Q_2< 0\,, \, u_2\leq -2u_1 \right \rbrace\,.
    \end{align*}
\end{minipage}      
\caption{Regions $(u_1,u_2) \in \mathcal{D}_2$ of application of the BKL maps  $\mathcal{B}_{\vec{e}}^{(g)}$ defined in Table \ref{tabla:kasnerexpgera}. We have defined $Q_1(u_1,u_2)=u_2^2 + u_1  u_2 + u_1^2 - u_1 - 2  u_2$ and $Q_2(u_1,u_2)=u_2^2+u_1^2+u_1 u_2+u_1-u_2$. Whenever two (or more) regions intersect, the result of applying any of the BKL maps associated to any of the regions is equal.}
\label{fig:regiones_grav_est}
\end{figure}

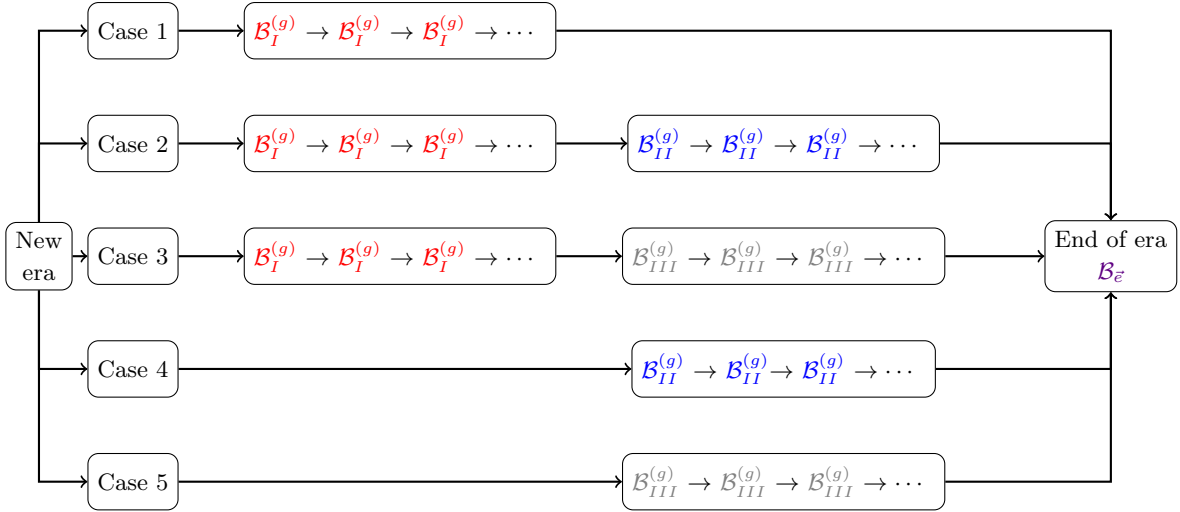
\begin{figure}
    \centering
\begin{tikzpicture}[
   node distance=1.8cm and 2.5cm, font=\footnotesize,
    box/.style={
        draw, rounded corners,
        minimum width=0.75cm,
        minimum height=0.75cm,
        align=center
    }
]

\node[box] (newera)  {New\\era};

\node[box, xshift=1.25cm, yshift=3cm] (case1) {Case 1};
\node[box, xshift=1.25cm, yshift=1.5cm] (case2) {Case 2};
\node[box, xshift=1.25cm, yshift=0cm] (case3) {Case 3};
\node[box, xshift=1.25cm, yshift=-1.5cm] (case4) {Case 4};
\node[box, xshift=1.25cm, yshift=-3cm] (case5) {Case 5};

\node[box, xshift=4.8cm, yshift=3cm] (k11) {
\begingroup \footnotesize
{\color{red} $\mathcal{B}_{I}^{(g)}$} $\rightarrow$ {\color{red}$\mathcal{B}_{I}^{(g)}$} $\rightarrow$ {\color{red}$\mathcal{B}_{I}^{(g)}$}  $\rightarrow \cdots$ \endgroup};
\node[box, xshift=4.8cm, yshift=1.5cm] (k21) {
\begingroup \footnotesize
{\color{red}$\mathcal{B}_{I}^{(g)}$} $\rightarrow$ {\color{red}$\mathcal{B}_{I}^{(g)}$} $\rightarrow$ {\color{red}$\mathcal{B}_{I}^{(g)}$}  $\rightarrow \cdots$ \endgroup};

\node[box, xshift=9.9cm, yshift=1.5cm] (k22) {\begingroup \footnotesize
{\color{blue}$\mathcal{B}_{II}^{(g)}$}  $\rightarrow$ {\color{blue}$\mathcal{B}_{II}^{(g)}$} $\rightarrow$ {\color{blue}$\mathcal{B}_{II}^{(g)}$} $\rightarrow \cdots$
\endgroup};

\node[box, xshift=4.8cm, yshift=0cm] (k31) {
\begingroup \footnotesize
{\color{red}$\mathcal{B}_{I}^{(g)}$} $\rightarrow$ {\color{red}$\mathcal{B}_{I}^{(g)}$} $\rightarrow$ {\color{red}$\mathcal{B}_{I}^{(g)}$}  $\rightarrow \cdots$ \endgroup};

\node[box, xshift=9.9cm, yshift=0cm] (k32) {\begingroup \footnotesize
{\color{gray}$\mathcal{B}_{III}^{(g)}$} $\rightarrow$ {\color{gray}$\mathcal{B}_{III}^{(g)}$} $\rightarrow$ {\color{gray}$\mathcal{B}_{III}^{(g)}$} $\rightarrow \cdots$
\endgroup};

\node[box, xshift=9.9cm, yshift=-1.5cm] (k4) {\begingroup \footnotesize
{\color{blue}$\mathcal{B}_{II}^{(g)}$} $\rightarrow$ {\color{blue}$\mathcal{B}_{II}^{(g)}$}$\rightarrow$ {\color{blue}$\mathcal{B}_{II}^{(g)}$} $\rightarrow \cdots$
\endgroup};

\node[box, xshift=9.9cm, yshift=-3cm] (k5) {\begingroup \footnotesize
{\color{gray}$\mathcal{B}_{III}^{(g)}$} $\rightarrow$ {\color{gray}$\mathcal{B}_{III}^{(g)}$} $\rightarrow$ {\color{gray}$\mathcal{B}_{III}^{(g)}$} $\rightarrow \cdots$
\endgroup};

\node[box,  xshift=14.25cm] (end) {End of era \\ \begingroup  \footnotesize
{\color{purple!80!black!95!red!60!blue}{$\mathcal{B}_{\vec{e}}$}}
\endgroup
};

\draw[->, thick]
    (newera.north) -- (0,3cm) -- (case1);
\draw[->, thick]
    (newera.north) -- (0,1.5cm) -- (case2);
\draw[->, thick] (newera) -- (case3);
\draw[->, thick]
    (newera.south) -- (0,-1.5cm) -- (case4);
\draw[->, thick]
    (newera.south) -- (0,-3cm) -- (case5);

\draw[->, thick] (case1) -- (k11);
\draw[->, thick] (case2) -- (k21);

\draw[->, thick] (k21) -- (k22);
\draw[->, thick] (case3) -- (k31);
\draw[->, thick] (case4) -- (k4);
\draw[->, thick] (case5) -- (k5);
\draw[->, thick] (k31) -- (k32);

\draw[->, thick] (k11) -- (14.25,3cm) -- (end.north);
\draw[->, thick] (k22) -- (14.25,1.5cm) -- (end.north);
\draw[->, thick] (k32) -- (end);
\draw[->, thick] (k4) -- (14.25,-1.5cm) -- (end.south);
\draw[->, thick] (k5) -- (14.25,-3cm) -- (end.south);

\end{tikzpicture}
\caption{Structure of Kasner eras with more than one epoch when gravitational walls are dominant. In Case 1, the era only features the Kasner season I until one the Kasner seasons $\vec{e}$ arises, implying the end of the era. In Case 2 (respectively, Case 3), the era starts with consecutive epochs in Kasner season I. At some point, the following epochs will all belong to Kasner seasons II (resp. III), until an epoch in one of the seasons $\vec{e}$ appears and the era finishes. Finally, in Case 4 (respectively 5) the era is characterized by successive epochs in Kasner season II (resp. III) until the era concludes with an epoch in one of the Kasner season $\vec{e}$. We have colored the various gravitational BKL maps to remark the different patterns of eras.}
\label{fig:ksdg}
\end{figure}

As a result, we have explicitly seen that Kasner seasons naturally appear in the BKL dynamics associated to gravitational walls in higher dimensions. Differently from the electric case, we have proven that, whenever epochs within a given era change of season, this season will continue all the way until the end of the era. Consequently, we learn that the nature of seasons will be different, according to the type of walls producing the subsequent BKL dynamics.

\subsubsection{Relation between electric and gravitational walls}

\begin{figure}[t!]
\centering
\begin{minipage}{0.49\linewidth}
    \includegraphics[scale=0.3]{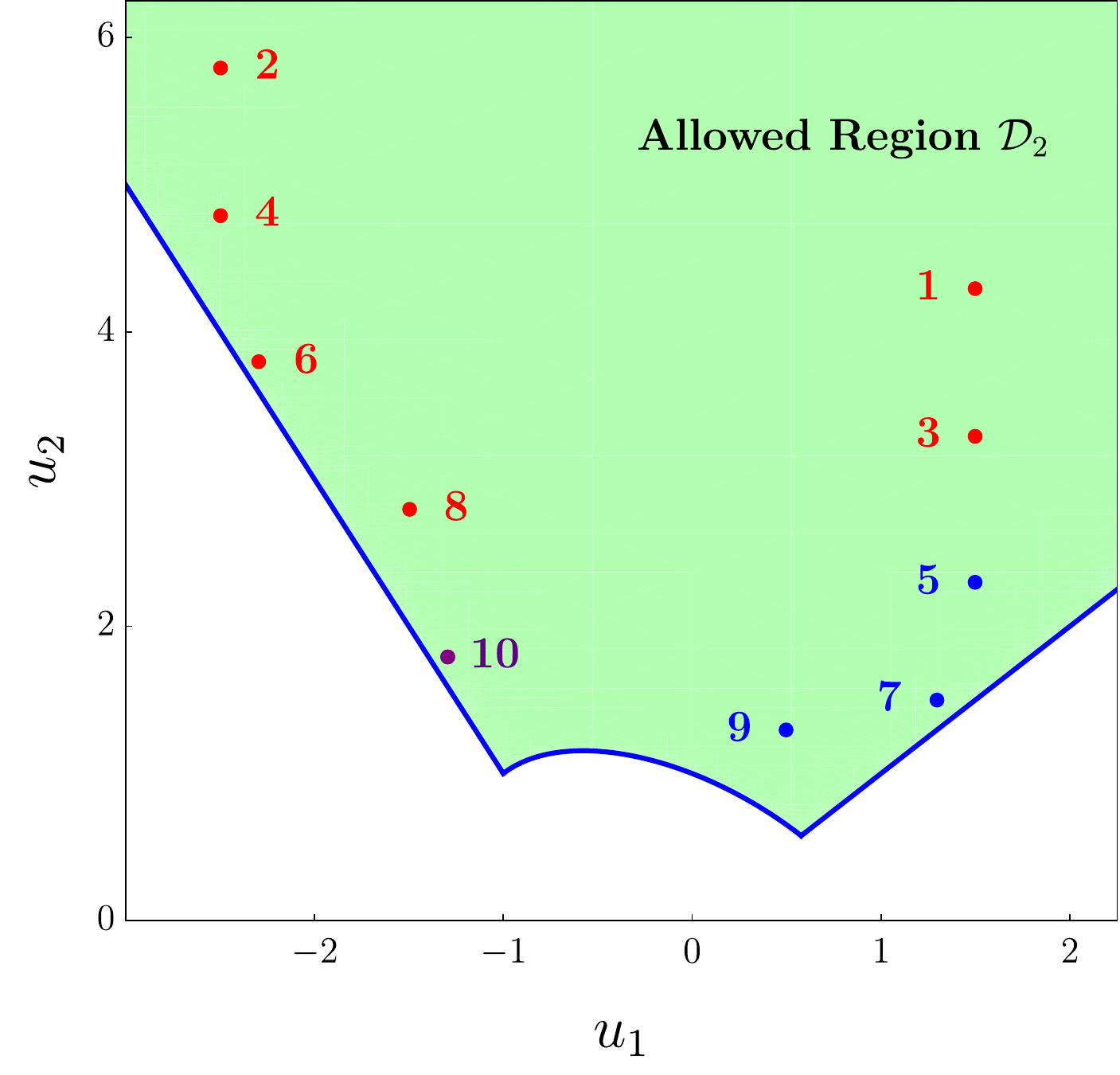}
\end{minipage}
\begin{minipage}{0.49\linewidth}
    \includegraphics[scale=0.3]{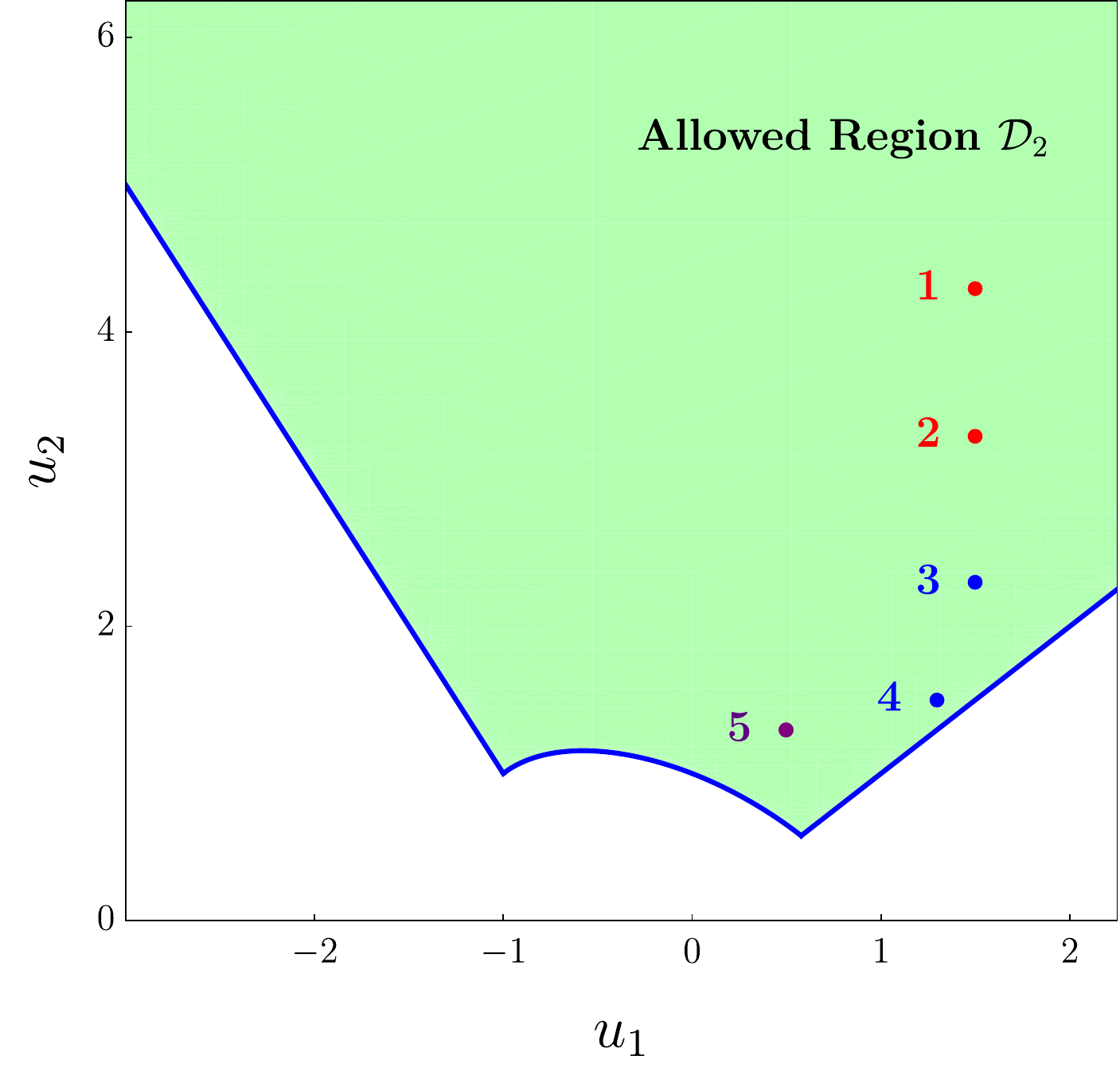}
\end{minipage}
\caption{\textbf{Left:} We present a given Kasner era in the interior of a five-dimensional AdS black hole of the form \eqref{eq:metricbh}. \textbf{Right:} We show a Kasner era associated with some Kasner dynamics driven by bounces against gravitational walls. We observe that the graphic on the left is formally equivalent to that on the right after \emph{deleting} the points at the leftmost part. }
\label{fig:kasnerseasons5}
\end{figure}

We conclude this section by pinpointing an intriguing relation between the electric BKL map $\mathcal{E}_5$ \eqref{eq:kassol5} and the gravitational BKL map $\mathcal{G}_5$ \eqref{eq:kasexpgrav}.  To this aim, let us assume we are within a certain Kasner epoch with exponents $p_j\geq p_k \geq p_l \geq p_m$. After a bounce against an electric wall, it is guaranteed that the least new Kasner exponent will be given by $p'_l$ --- see \eqref{eq:kassol5}. As a consequence, if we now compute the Kasner exponents $ \left\lbrace p''_j,p''_k,p''_l,p''_m \right\rbrace$ after a second bounce towards another electric wall:
\begin{align}
\notag
\left\lbrace p''_j,p''_k,p''_l,p''_m\right\rbrace&=\mathcal{E}_5^2 \left ( \left \lbrace p_j,p_k,p_l,p_m \right\rbrace \right)\\ &= \left\lbrace
   \frac{1-p_k+p_m}{\Delta_p}\,, \, \frac{1-p_j+p_m}{\Delta_p}\,, \,\frac{p_j+p_k-1}{\Delta_p}\,,\, \frac{p_l}{\Delta_p}\right\rbrace\,, 
   \label{eq:kasexp2ele}
\end{align}
where $\Delta_p=2-p_j-p_k+p_m$. We observe that \eqref{eq:kasexp2ele} and that \eqref{eq:kasexpgrav} are formally equivalent, after the permutation of $p''_l$ with $p''_m$. As a matter of fact, the BKL laws associated to gravitational walls are obtained by composing the electric BKL laws. This follows directly by
\begin{equation}
 \mathcal{K}_I^{(g)}=\mathcal{K}_I \circ \mathcal{K}_I\,, \quad \mathcal{K}_{II}^{(g)}=\mathcal{K}_I \circ \mathcal{K}_{II}\,, \quad \mathcal{K}_{III}^{(g)}=\mathcal{K}_{II} \circ \mathcal{K}_I\,,
\end{equation}
and by the results presented in Table \ref{tabla:kasnerexpgera}. Consequently, while in four dimensions there is no distinction at all between gravitational and electric walls, in $D=5$ we obtain the intriguing relation
\begin{equation}
    \mathrm{Electric}\,\,  \mathrm{walls} \sim \sqrt{\mathrm{Gravitational} \, \, \mathrm{walls}}\,.
    \label{eq:doublecopy}
\end{equation}
We illustrate the implications of this result in the space $(u_1,u_2) \in \mathcal{D}_2$ in Figure \ref{fig:kasnerseasons5}. We close this section by noting  that \eqref{eq:doublecopy} is highly reminiscent of the double copy formalism \cite{Bern:2008qj,Bern:2010ue,Saotome:2012vy,Monteiro:2014cda}. 

\subsection{Kasner seasons in $D >5$}

Let us now examine the BKL dynamics arising in the interior of charged AdS $D$-dimensional black holes of the form \eqref{eq:ans} with $D >5$, identifying the pattern of seasons in general higher space-time dimensions. The collision law $\mathcal{E}_{D}$ for the Kasner exponents $\{p_{j_0},p_{j_1},\dots p_{j_m}\}$ with $m=D-2$ was already derived in \eqref{eq:kassol}. If $p_{j_m}$  stands for the least Kasner exponent, one has
\begin{equation}
    \mathcal{E}_D: (p_{j_0}\,,\dots \,,p_{j_m}) \rightarrow \left ( \frac{p_{j_0}+a\, p_{j_m}}{1+a \,p_{j_m}}\,, \dots\,,\frac{p_{j_{m-1}}+a\, p_{j_m}}{1+a\, p_{j_m}}\,, -\frac{p_{j_m}}{1+a \,p_{j_m}} \right)\,,
\end{equation}
with $a=\frac{2}{D-3}$. Assume the exponents before the bounce are ordered as $p_{j_0} >p_{j_1} > \dots >p_{j_m}$ (we will consider the general case in which no exponents coincide. We refer the reader to Section \ref{subsec:restricted_triangle}, where this is partly considered). As argued in the appendix, there can be at most $D-3$ positive Kasner exponents, implying that\footnote{The case $p_{j_m}=0$ forces all exponents being zero except for $p_{j_1}=1$.} $p_{j_m}<0$.   Using the parametrization \eqref{eq:parkas}, there exist parameters $\vec{u}=(u_1,\dots,u_{D-3}) \in \mathcal{D}_{D-3}$ as defined in \eqref{eq:dregion} such that $p_{j_0}=\mathfrak{p}_{D-2}(\vec{u})$, $p_{j_1}=\mathfrak{p}_{D-3}(\vec{u}),\dots \,$, $p_{j_m}=\mathfrak{p}_0 (\vec{u})$. The resulting Kasner exponents after the bounce will be given by new parameters $(u_1', \dots, u_{D-3}')\in \mathcal{D}_{D-3}$ and a new reordering $p'_{\sigma_{A(j_0)}}>p'_{\sigma_{A(j_1)}}> \dots > p'_{\sigma_{A(j_m)}}$, where $\sigma_A: \mathbb{Z}_{D-1} \rightarrow \mathbb{Z}_{D-1}$ is a permutation and $A$ belongs to an index set. As before, we decompose the collision law $\mathcal{E}_D$ into a BKL law $\mathcal{K}_A$ and a BKL reordering\footnote{For the sake of convenience, in general higher space-time dimensions we use arabic numerals for the indices in $A$.} $\sigma_A$:
\begin{table}[t!]
\centering
\renewcommand{\arraystretch}{2}

\begin{tabular}{|c|c|c|c|}
\cline{1-4}
\textbf{BKL map} $\mathcal{B}_A$ & $\mathcal{B}_1$ &  $\mathcal{B}_{l+1}\,,$ $D-4 \geq l \geq1$ & $\mathcal{B}_{D-2}$ \\ 
\hline
{\textbf{Range of}} & $1+\gamma< \tilde{u}_{D-3}\,$ and & $1+\gamma> \tilde{u}_{D-3}\,,$ & \multirow{3}{*}{$\lambda-1< \tilde{u}_{D-3}$}\\
& & $-u_{l} > -\tilde{u}_{D-3} > -{u}_{l+1}\,,$ &  \\
{\textbf{Application}} & $\lambda-1> \tilde{u}_{D-3}$ & and $\, \lambda-1> \tilde{u}_{D-3}$ &  \\
\hline
\textbf{BKL law} & $u_1'=\tilde{u}_{D-3}-\gamma-1$ & $u_j'=u_j+\tilde{u}_{D-3}\,,$ $\, j \leq l$ & $u_1'=\dfrac{\tilde{u}_{D-3}-1-\gamma}{\lambda-\tilde{u}_{D-3}}$ \\ \multirow{2}{*}{$\mathcal{K}_A(\vec{u})=\vec{u}\,'$} & $u_k'=u_{k-1}+\tilde{u}_{D-3}$& $u_{l+1}'=-u_{D-3}$ & $u_k'=\dfrac{u_{k-1}+\tilde{u}_{D-3}}{\lambda-\tilde{u}_{D-3}}$ \\ & $k \in \{2, \dots, D-3\}$ & $u'_{j+1}=u_{j}+\tilde{u}_{D-3}\,,$ $\, j>l$ & $k \in \{2, \dots, D-3\}$ \\
\hline
\textbf{BKL reordering}  & \multirow{2}{*}{$\{j_0,j_m,j_1,j_2, \dots \}$}& $\{j_0,j_1, \dots,j_{l},$ & \multirow{2}{*}{$\{j_m,j_0,j_1, \dots \}$} \\ $p_{\sigma(J)}'=\mathfrak{p}_{N}(\vec{u}\,')$ & & $j_m,j_{l+1},j_{l+2},\dots\}$ &  \\
\hline
\end{tabular}
\caption{New Kasner exponents $
p'_J$ in terms of the initial ones $
p_J$, where $J=\{j_0,j_1,\dots, j_m\}$ and $N=\{D-2,D-3,\dots,1,0\}$. It is assumed that $p_{j_0} >p_{j_1}>\dots >p_{j_m}$. Note that the different reorderings indicate the new ordering of the new exponents, from larger to smaller. We defined $\tilde{u}_i=(a+1)u_i$, while $\lambda$ were $\gamma$ were given in \eqref{eq:deflg}. If $\vec{u} \in \mathcal{D}_{D-3}$, then $\vec{u}\,' \in \mathcal{D}_{D-3}$ for all BKL maps. Also, for $\mathcal{B}_{D-3}$, the second condition in \emph{Range of Application} would just read $-u_{D-4} \geq -\tilde{u}_{D-3}$. }

\label{tabla:kasnerexpD}
\end{table}
\begin{equation}
    \mathcal{E}_D: \{p_{j_1},\dots, p_{j_m}\} \rightarrow \left\lbrace p_{j_1}',\dots,p_{j_m}' \right\rbrace\longleftrightarrow \left\lbrace \begin{matrix}
        \mathcal{K}_A: \mathcal{D}_{D-3} \rightarrow \mathcal{D}_{D-3}\\
        \sigma_A: \mathbb{Z}_{D-1} \rightarrow \mathbb{Z}_{D-1}
    \end{matrix} \right.\,, \quad A={1,2, \dots, D-2}\,.
\end{equation}
The fact that $A$ runs over $(D-2)$ different values is seen from the fact that $p'_{j_0} >p'_{j_1}>\dots >p'_{j_{m-1}}$ and that $p'_{j_m}>0>p'_{j_{m-1}}$. If we define the composite functions $\mathcal{B}_A:(\mathcal{K}_A, \sigma_A):\mathcal{D}_{D-3} \times \mathbb{Z}_{D-1} \rightarrow \mathcal{D}_{D-3} \times \mathbb{Z}_{D-1}$ for $A=1,2,\dots, D-2$, we have that:
\begin{itemize}
    \item Suppose\footnote{Observe that $1+a p_{j_m} >0$, as it may be learnt from the minimum value for $p_{j_m}$ (see appendix).} $p_{j_0}+a p_{j_m}>-p_{j_m} >p_{j_1}+a p_{j_m}$. This translates into the conditions $\lambda(\vec{u})-1-(a+1) u_{D-3}>0$ and $1-a u_{D-3}+\sum_{i=1}^{D-4} u_i<0$. In this case, it turns out that:
    \begin{align}
       & \mathcal{K}_1(\vec{u})=\left (a u_{D-3}-\gamma-1 ,u_1+a u_{D-3},u_2+ au_{D-3}, \dots, u_{D-4}+a u_{D-3}\right)\,, \\
       & \sigma_1\left (\{j_1,\dots, j_m\}\right)=\{j_0,j_{m},j_1, \dots, j_{m-1}\}\,,
    \end{align}
 where $\gamma=\sum_{j=1}^{D-3} u_j$ as in \eqref{eq:deflg}.
     \item Assume that $p_{j_0}+a p_{j_m}>\dots > p_{j_l}+a p_{j_m}>-p_{j_m} >p_{j_{l+1}}+a p_{j_{l+1}}$, with $D-4 \geq l \geq 1$. This translates into the conditions $\lambda(\vec{u})-1-(a+1) u_{D-3}>0$, $1-a u_{D-3}+\sum_{i=1}^{D-4} u_i>0$, $-u_1 > -u_2 > \dots > -u_{l} >(a+1)u_{D-3} >-u_{l+1}>-u_{l+2}>\dots$. In this case, it turns out that:
    \begin{align}
       & \mathcal{K}_{l+1}(\vec{u})=\left (u_1 +a u_{D-3}, \dots, u_{l}+a u_{D-3}, -u_{D-3}, u_{l+2}+a u_{D-3}, \dots \right)\,, \\
       & \sigma_{l+1}\left (\{j_0,\dots, j_m\}\right)=\{j_0,j_1, \dots, j_{l},j_{m}, j_{l+1}, \dots, j_{m-1}\}\,,  \quad l=1,\dots D-4\,.
    \end{align}

    \item Suppose instead that $-p_{j_m} > p_{j_0}+a p_{j_m}$. This implies that $\lambda(\vec{u})-1-(a+1) u_{D-3}<0$. The subsequent BKL law $\mathcal{K}_{D-2}$ and reordering $\sigma_{D-2}$ are given by:
    \begin{align}
       & \hspace{-0.2cm} \mathcal{K}_{D-2}(\vec{u})=\frac{1}{\lambda(\vec{u})-(a+1) u_{D-3}}\left (au_{D-3} -\gamma-1, u_1+a u_{D-3},u_2 +a u_{D-3},\dots \right)\,, \\
       & \hspace{-0.2cm}  \sigma_{D-2}\left (\{j_0,\dots, j_m\}\right)=\{j_m,j_0, j_1,j_2 \dots, j_{m-1}\}\,.
    \end{align}

\end{itemize}

Tha action of the various BKL maps $\left\lbrace \mathcal{B}_A \right\rbrace_{A=1}^{D-2}$ is summarised in Table \ref{tabla:kasnerexpD}. On the one hand, we observe that the first $D-3$ BKL maps $\left\lbrace \mathcal{B}_A \right\rbrace_{A=1}^{D-3}$ preserve the direction of the largest Kasner exponent. As a result, these correspond to transitions between epochs belonging to the same era. However, $\mathcal{B}_{D-2}$ does modify the direction of the largest exponent: as such, it represents a change of era. Therefore, we identify $D-2$ Kasner seasons in general space-time dimensions $D \geq 4$, where $D-3$ seasons correspond to transitions between epochs of the same era, while only one season is associated with the change of era. Consequently, Kasner seasons are a natural fetaure in general higher space-time dimensions and their structure and variety get richer as $D$ increases.

\subsection{Restricted motion on a triangular domain}
\label{subsec:restricted_triangle}

In this section, we would like to focus on a symmetry-enhanced subsector in which \((D-3)\) of the spatial Kasner exponents coincide,
\begin{equation}
p_2=p_3=\cdots=p_{D-2}\equiv p_\perp\, .
\end{equation}
This restriction reduces the independent Kasner exponents to the triple \((p_0,p_1,p_\perp)\), which must satisfy the (reduced) Kasner constraints
\begin{align}
p_0+p_1+(D-3)p_\perp &= 1\,, \label{eq:kasner_reduced_sum}\\
p_0^2+p_1^2+(D-3)p_\perp^2 &= 1\,. \label{eq:kasner_reduced_sq}
\end{align}
Geometrically, the motion in configuration space that is generically confined to the interior of a regular \((D-2)\)-simplex becomes confined to a two-dimensional slice: the equal-exponent condition forces the trajectory to lie on a triangular domain inside that simplex.

In terms of the configuration-space variables \(f_i(\rho)\), the equality of the \((D-3)\) exponents implies that the corresponding \((D-4)\) velocities vanish identically. Using the relation between Kasner exponents and velocities (cf.\ \eqref{eq:kasnerxp}), one finds
\begin{equation}
v_2=v_3=\cdots=v_{D-3}=0\, .
\end{equation}
Consequently, the associated coordinates are constants of motion during the asymptotic geodesic evolution, and we may, without loss of generality, choose the origin of the \(\{f_i\}\) so that
\begin{equation}
f_2=f_3=\cdots=f_{D-3}=0\, .
\end{equation}
From the bulk perspective, this reduction corresponds to an enhanced spatial isometry among the \((D-4)\) directions labeled by \(i=2,\dots,D-3\). Consistency of the ansatz with this enhanced symmetry requires that the corresponding \((D-4)\) massive vector fields be identified,
\begin{equation}
\phi_2=\phi_3=\cdots=\phi_{D-3}\equiv\phi_\perp\, .
\end{equation}

With these constraints implemented, only three independent exponential walls remain relevant in the late-interior regime: the motion is bounded by the three hyperplanes
\begin{equation}
\text{Wall }{\rm W}_m:\;
-(D-2-m)a_m\,\frac{f_m}{\rho}
+\sum_{q=0}^{m-1} a_j\,\frac{f_q}{\rho}
=1\,,
\qquad m=0,\,1,\,\perp\,,
\label{eq:restricted_walls_general}
\end{equation}
which define a triangular billiard domain in the plane spanned by \((f_0/\rho,f_1/\rho)\). Writing these three walls explicitly gives
\begin{align}
{\rm W}_0:\qquad &\frac{f_0}{\rho}=-\frac{1}{D-2}\,, \label{eq:W0_restricted}\\
{\rm W}_1:\qquad &\frac{f_0}{\rho}=1+\sqrt{(D-1)(D-3)}\,\frac{f_1}{\rho}\,, \label{eq:W1_restricted}\\
{\rm W}_{\perp}:\qquad &\frac{f_0}{\rho}=1-\sqrt{\frac{D-1}{D-3}}\,\frac{f_1}{\rho}\,. \label{eq:WDm2_restricted}
\end{align}
\begin{figure}[t!]
    \centering
    \includegraphics[width=0.40\linewidth]{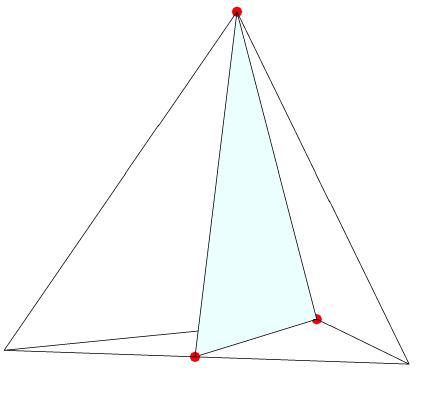}
    \includegraphics[width=0.5\linewidth]{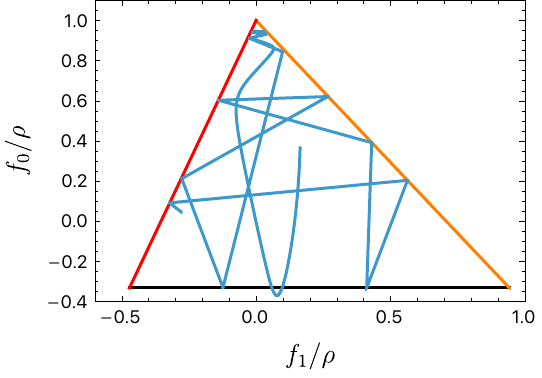}
    \caption{\textbf{Left:} Two dimensional slicing of the tetrahedron in $D=5$ when $p_2=p_3\equiv p_{\perp}$, \textbf{Right:} Late time of motion of a particle restricted on that triangle.}
    \label{fig:placeholder1}
\end{figure}
This triangular domain is generically \emph{not} equilateral for \(D\ge 5\): the two slanted walls \eqref{eq:W1_restricted} and \eqref{eq:WDm2_restricted} have unequal slopes, reflecting the fact that the symmetry reduction selects a non-regular two-dimensional slice of the higher-dimensional regular simplex. This should be contrasted with the \(D=4\) case, where the full billiard already lives in two dimensions and the late-time motion is confined to an equilateral triangle.

We will now study the bouncing rule from each wall. These can be obtained by solving the (\ref{bouncerule}),
\begin{equation}
    \frac{\dot{f}_1-\lambda_1^{(m)}}{\dot{f}_1^{(0)}-\lambda_1^{(m)}}=\frac{\dot{f}_0-\lambda_0^{(m)}}{\dot{f}_0^{(0)}-\lambda_0^{(m)}}\,, \quad  m=0,1, \perp\,,
    \label{eq:rebotesim}
\end{equation}
where $\left\lbrace \dot{f}_0^{(0)}, \dot{f}_1^{(0)} \right\rbrace $ denote the values of the derivatives just before the bounce, while $\left\lbrace \dot{f}_0, \dot{f}_1  \right\rbrace $ stand for the aftermath derivatives. Note that these must satisfy \eqref{eq:sumlig}, so that 
\begin{equation}
    \left (\dot{f}_0^{(0)} \right) ^2+ \left (\dot{f}_1^{(0)} \right) ^2=\dot{f}_0^2+ \dot{f}_1^2 =1\,.
    \label{eq:simlig}
\end{equation}
For each wall, \eqref{eq:rebotesim} and \eqref{eq:simlig} conform a quadratic system of two equations for $\{\dot{f}_0, \dot{f}_1\}$. One solution corresponds to the trivial case $\dot{f}_0=\dot{f}_0^{(0)}$ and $\dot{f}_1=\dot{f}_1^{(0)}$. The second one represents the physical solution for a bounce against the $m$-th wall with $m=0,1,\perp$. In any of these cases, this bouncing rule may be conveniently expressed in terms of the associated Kasner exponents, although it takes a different form depending on if $m=n$ with $n=0,1$ or $m=\perp$. To see this, let $\{p_0,p_1,p_{\perp} \}$ be the post-collision Kasner exponents and $\{p_0^{(0)},p_1^{(0)},p_{\perp}^{(0)} \}$ be the pre-collision Kasner exponents. For a bounce onto the $n$-th wall ($n=0,1$), one has:
\begin{align}
    \frac{p_i}{p_n}&=-\frac{2}{D-3}-\frac{p^{(0)}_{i}}{p^{(0)}_{n}},\,\quad\,\text{with}\,\quad\, i=0,1,\perp,  \quad n=0,1 \, \quad \text{and}\, \quad i\neq n\,,
\end{align}
This can be solved exactly, finding that
\begin{align}
    p_i&=\frac{p_i^{(0)}+a p_n^{(0)}}{1+a p_n^{(0)}},\qquad\text{and}\,\quad\, p_n=-\frac{p_n^{(0)}}{1+a p_n^{(0)}},\quad a=\frac{2}{D-3}\,. 
\end{align}
On the other hand, for the $\perp$ wall, the bouncing rule becomes
\begin{equation}
    \frac{p_i}{p_\perp}=-2(D-3)-\frac{p^{(0)}_{i}}{p^{(0)}_{\perp}},\,\quad\,\text{with}\,\quad\, i=0,1\,.
\end{equation}
\begin{figure}
    \centering
    \includegraphics[width=0.6\linewidth]{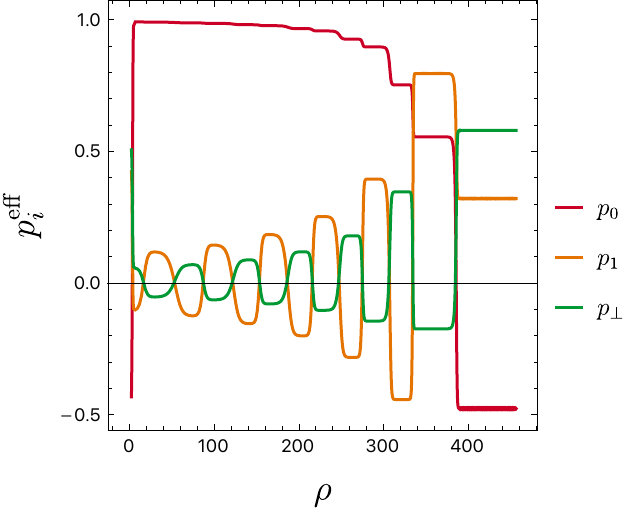}
    \caption{Kasner exponents $(p_0,p_1,p_\perp)$ for the triangular billiard in $D=5$. These exponents satisfy $p_0+p_1+2p_\perp=1$ and $p_0^2+p_1^2+2p_\perp^2=1$.}
    \label{fig:placeholder2}
\end{figure}
This can be solved to get the Kasner exponents after a bounce against the $\perp$ wall:
\begin{align}
    p_i&=\frac{p_i^{(0)}+b p_\perp^{(0)}}{1+b p_\perp^{(0)}},\qquad\text{and}\,\quad\, p_\perp=-\frac{p_\perp^{(0)}}{1+b p_\perp^{(0)}},\quad b=2(D-3)\,.
\end{align}
It would be interesting to map these chaotic bounces onto some fundamental domain. We leave such a study for future exploration.

\subsubsection{Numerical method for $D=5$}
In this section we study the numerical solutions in \(D=5\) in detail. Upon substituting the metric ansatz into the Einstein equations together with the matter-field equations of motion in (\ref{eq: Einsteinplusmatter}), the system reduces to eight independent\footnote{In any spacetime dimensions $D$, there are $2(D-1)$ independent equations.} ordinary differential equations for the radial fields. These equations are supplemented by the appropriate constraint relations and boundary conditions at the AdS boundary and at the horizon, which we use to construct the black hole solutions and to follow their continuation into the interior,
\begin{align}
&0= -3 F(z) f_1'(z)^2 - z^2 F(z) e^{-f_1(z)-f_2(z)}\phi_3'(z)^2 - z^2 F(z) e^{f_2(z)-f_1(z)}\phi_2'(z)^2 \nonumber \\
&\quad - z^2 F(z) e^{2 f_1(z)} \phi_1'(z)^2 - F(z) f_2'(z)^2 + \frac{6F(z)H'(z)}{z} - \frac{\mu_0^2 e^{2 H(z)} \phi_0(z)^2}{F(z)},\\
&0= -\frac{6 f'(z)}{z} + \frac{6 F(z) H'(z)}{z} + \frac{24 F(z)}{z^2} + \mu_2^2 \phi_2(z)^2 e^{f_2(z)-f_1(z)} + \mu_3^2 \phi_3(z)^2 e^{-f_1(z)-f_2(z)} \nonumber \\
&\quad + \mu_1^2 e^{2 f_1(z)} \phi_1(z)^2 + z^2 e^{2 H(z)} \phi_0'(z)^2 - \frac{20}{z^2}, \\
&0= 6 z F(z) e^{f_1(z)+f_2(z)} f_2''(z) + 6 F(z) e^{f_1(z)+f_2(z)} f_2'(z) - 3 z^3 F(z) e^{2f_2(z)} \phi_2'(z)^2 \nonumber \\
&\quad + 3 z^3 F(z) \phi_3'(z)^2 + z^4 f_2'(z) \phi_0'(z)^2 e^{f_1(z)+f_2(z)+2 H(z)} + \mu_1^2 z^2 \phi_1(z)^2 e^{3 f_1(z)+f_2(z)} f_2'(z) \nonumber \\
&\quad - 20 e^{f_1(z)+f_2(z)} f_2'(z) + \mu_2^2 z e^{2f_2(z)} \phi_2(z)^2 (z f_2'(z) - 3) + \mu_3^2 z \phi_3(z)^2 (z f_2'(z) + 3), \\
&0= 6 z F(z) e^{f_1(z)+f_2(z)} f_1''(z) + 6 F(z) e^{f_1(z)+f_2(z)} f_1'(z) - 2 z^3 F(z) e^{3f_1(z)+f_2(z)} \phi_1'(z)^2 \nonumber \\
&\quad + z^3 F(z) e^{2 f_2(z)} \phi_2'(z)^2 + z^3 F(z) \phi_3'(z)^2 + z^4 f_1'(z) \phi_0'(z)^2 e^{f_1(z)+f_2(z)+2 H(z)} \nonumber \\
&\quad + \mu_1^2 z \phi_1(z)^2 e^{3 f_1(z)+f_2(z)} (z f_1'(z) - 2) + \mu_2^2 z^2 e^{2 f_2(z)} \phi_2(z)^2 f_1'(z) \nonumber \\
&\quad - 20 e^{f_1(z)+f_2(z)} f_1'(z) + \mu_3^2 z \phi_3(z)^2 (z f_1'(z) + 1) + \mu_2^2 z e^{2 f_2(z)} \phi_1(z)^2,\\
&0= z F(z) \left( (1 - z H'(z)) \phi_0'(z) - z \phi_0''(z) \right) + \mu_0^2 \phi_0(z),\\
&0= z \left( \phi_1'(z) \left( z f'(z) + F(z)(2 z f_1'(z) - z H'(z) - 1) \right) + z F(z) \phi_1''(z) \right) - \mu_1^2 \phi_1(z), \\
&0= z \left( z F(z) \phi_2''(z) - \phi_2'(z) \left( -z f'(z) + z F(z)(f_1'(z) - f_2'(z) + H'(z)) + F(z) \right) \right) - \mu_2^2 \phi_2(z), \\
&0= z \left( z F(z) \phi_3''(z) - \phi_3'(z) \left( -z f'(z) + z F(z)(f_1'(z) + f_2'(z) + H'(z)) + F(z) \right) \right) - \mu_3^2 \phi_3(z)
\end{align}
We solve this coupled boundary-value problem numerically using a shooting method. Concretely, we expand the eight fields $X(z)=\{F(z),H(z),\phi_0(z),\phi_1(z),\phi_2(z),\phi_3(z),f_1(z),f_2(z)\}$ in a regular Taylor series about the horizon location \(z=z_h\),
\begin{eqnarray}\label{eq:horizon_expansion_general}
    X(z)&=&\sum_{n=0}^{N}X_n(z_h)\,(z-z_h)^n+\mathcal{O}(z-z_h)^{N+1}
\end{eqnarray}
with the horizon located at \(z=z_h\), defined by the simple zero of the blackening factor,
\begin{equation}
F(z_h)=0\,, \qquad F'(z_h)\neq 0\,,
\end{equation}
and with regularity imposed by requiring that all remaining fields admit a smooth Taylor expansion about \(z_h\) and that curvature invariants remain finite there. Substituting the truncated series ansatz \eqref{eq:horizon_expansion_general} into the equations of motion and solving order by order in \((z-z_h)\) fixes the higher-order coefficients \(\{X_{n\ge 1}\}\) algebraically in terms of a finite set of independent horizon data, after using the residual gauge freedom to eliminate redundant parameters. 

In practice we terminate the expansion at some finite order \(N\ge 1\) and use the resulting expressions to generate initial conditions at \(z=z_h+\varepsilon\) with \(\varepsilon\ll 1\). We then integrate the coupled first-order system outward to the AdS boundary and tune the independent horizon data (the shooting parameters) so that the solution satisfies the desired asymptotic boundary conditions. An analogous integration inward from \(z=z_h-\varepsilon\) provides the continuation across the horizon and into the black hole interior. 

Upon substituting the near-horizon expansions into the equations of motion and enforcing regularity at each order, we find the $X_n(z_h)$ in terms of 
their horizon values and the first derivative of the blackening factor \(F'(z_h)\). Regularity additionally fixes the temporal gauge potential to vanish at the horizon,
\begin{equation}
\phi_0(z_h)=0\,,
\end{equation}
so that the corresponding one-form is smooth in a regular (Euclidean) continuation. We construct the near-horizon initial data by solving the equations order-by-order in a Taylor expansion about \(z_h\), keeping terms up to fifth order, i.e.\ truncating the series at \(N=5\).

At the asymptotic AdS boundary the field behaves as,
\begin{equation}
    F\xrightarrow[z\to 0]{}\frac{1}{L^2_{\text{AdS}}},\qquad H, f_i\xrightarrow[z\to 0]{} 0,\qquad 
\end{equation}
and for the massive vector fields, the boundary components admit the asymptotic expansion
\begin{equation}
\phi_i(z,x)\xrightarrow[z\to 0]{}
\phi^{(0)}_i(x)\,z^{\alpha}+\phi^{(1)}_i(x)\,z^{\beta}+\cdots,
\qquad 
\alpha=D-\Delta-2,\quad \beta=\Delta-1,
\end{equation}
with
\begin{equation}
\Delta=\frac12\left(D-1+\sqrt{(D-3)^2+4m^2L_{\text{AdS}}^2}\right).
\end{equation}
These massive vector fields must satisfy the Breitenlohner--Freedman (BF) stability bound in AdS,
\begin{equation}
m^2 L_{\text{AdS}}^2 \ge -\frac{(D-3)^2}{4}.
\end{equation}
which ensures that the near-boundary falloff exponents are real.

\section{Holographic probe to the epochs and eras: thermal $a$-function}\label{sec:holographic_probe}

A central question in holographic studies of black hole interiors is how to access the behind-horizon geometry from boundary data. A standard class of probes consists of correlators of heavy operators, which in the semiclassical limit are controlled by (spacelike)
bulk geodesics anchored at the AdS boundary \cite{Fidkowski:2003nf,Festuccia:2005pi,Frenkel:2020ysx}. Complementary information can also be extracted from intrinsically quantum-information observables, such as entanglement measures
and notions of complexity, which have been argued to capture certain aspects of the interior physics.

For instance, in the geodesic approximation for heavy operators, equal-time two-point functions are
controlled by spacelike bulk geodesics anchored at the AdS boundary. For a static,
translationally invariant metric of the form
\begin{equation}
ds^2 = g_{tt}(r)\,dt^2 + g_{rr}(r)\,dr^2 + g_{xx}(r)\,d\vec{x}^{\,2}\,,\qquad g_{tt}(r)<0\ \text{outside the horizon,}
\label{eq:static_metric_geodesic}
\end{equation}
spacelike geodesics admit conserved quantities associated with the Killing symmetries, in
particular an ``energy'' $E\equiv -g_{tt}\,\dot t$ and a spatial momentum $p\equiv g_{xx}\,\dot x$
(dots denote derivatives with respect to an affine parameter). Imposing the spacelike
normalization condition $g_{\mu\nu}\dot x^{\mu}\dot x^{\nu}=+1$ yields a radial equation of the
schematic form
\begin{equation}
g_{rr}(r)\,\dot r^{\,2} = 1+\frac{E^2}{g_{tt}(r)}-\frac{p^2}{g_{xx}(r)} \equiv V_{\rm eff}(r;E,p)\,,
\label{eq:Veff_geodesic}
\end{equation}
with an allowed region determined by $V_{\rm eff}(r;E,p)\ge 0$, and turning points
given by $V_{\rm eff}=0$.

In our model, the massive vector fields backreact on the geometry in such a way that the
would-be second horizon is destroyed, and correspondingly
$g_{tt}(r)$ develops a local maximum at some radius $r=r_\star$. This feature creates a
potential barrier for spacelike trajectories: because the $E^2/g_{tt}(r)$ term in
\eqref{eq:Veff_geodesic} is most restrictive precisely where $g_{tt}(r)$ is extremal, a local
maximum of $g_{tt}$ forces $V_{\rm eff}(r;E,p)$ to change sign before the geodesic can reach
deeper radii. Equivalently, as one attempts to increase the boundary separation (which
requires pushing the turning point inward), the turning point asymptotes to $r_\star$ and
cannot be continued past it: the geodesic becomes stuck at $r_\star$ rather than penetrating
into the excised region. This type of obstruction prevents neutral geodesics from accessing epochs in which the effective time-direction Kasner exponent satisfies $p_0>0$ \cite{Hartnoll:2020rwq}.

As a consequence, the usual geodesic prescription for boundary two-point functions fails to access the deep-interior physics close to the black hole singularity. One way around the
barrier is to consider \emph{charged} geodesics, which can probe past $r_\star$ and reach arbitrarily close to the singularity in the presence of bulk gauge fields~\cite{Carballo:2024hem}.
However, this mechanism is not universal: it relies on coupling the probe to gauge fields and
therefore does not apply to interiors whose near-singularity dynamics is driven purely by
gravitational walls (in which case the charged and neutral geodesics coincide).

Quantum-information observables provide alternative windows into the interior. For instance,
extremal surfaces computing entanglement entropy typically develop their own turning points
and, in generic static geometries, do not extend all the way to the spacelike singularity.
On the other hand, certain notions of holographic complexity have been argued to probe arbitrarily
deep regions of the interior and, in particular, can act as singularity diagnostics for a class of ``complexity=anything'' proposals~\cite{Jorstad:2023kmq,Arean:2024pzo}. We leave a systematic study of complexity-based probes of BKL epochs and eras for future work.

In this section, we will focus on a simple holographic diagnostic that is not obstructed by the turning-point barrier and remains well-defined throughout the interior. This probe
is encoded in a thermal $a$-function $a_T$ that is monotonically decreasing along the inward radial direction \cite{Caceres:2022smh}, thereby providing an unambiguous diagnostic of the deep-interior regime.

\subsection{Thermal $a$-function and interior epochs}

In this section, we explain how a candidate \emph{thermal $a$-function} can be used to analyze the entire radial evolution of an asymptotically AdS black hole geometry---from the AdS boundary, through the horizon, and deep into the interior all the way to the spacelike singularity. This viewpoint has been exhaustively explored in \cite{Frenkel:2020ysx,Caceres:2022smh,Caceres:2022hei,Caceres:2023zft,Arean:2024pzo,Carballo:2024hem,Caceres:2023mqz,Caceres:2024edr} and known as ``trans-IR flow'' RG flow into the black hole interiors, where continuing the holographic RG flow beyond the IR region naturally parametrizes behind-horizon physics and admits a monotonic thermal generalization of holographic $a$-functions under the null energy condition (NEC).

To construct such a function, we impose the NEC
\begin{equation}
    T_{\mu\nu}\,\kappa^{\mu}\kappa^{\nu}\geq 0\,,
\end{equation}
where $T_{\mu\nu}$ is the bulk stress-energy tensor and $\kappa^{\mu}$ is any null vector. In holography, the bulk radial direction is identified with the field-theory energy scale, so we choose a radially directed null vector. For the metric ansatz in Eq.~(\ref{eq:metricbh}), a convenient choice is
\begin{equation}
    \kappa^{\mu}= \frac{e^{H(z)}}{F(z)}\delta^\mu_{t}+\delta^{\mu}_{z}\,,
\end{equation}
which is manifestly null. Using the metric in Eq.~(\ref{eq:metricbh}), the NEC along this congruence becomes
\begin{eqnarray}
    T_{\mu\nu}\kappa^{\mu}\kappa^{\nu}
    &=& \frac{D-2}{z}H'(z)-\frac{(D-1)(D-2)}{4}\sum_{j=1}^{D-3}f_{j}'(z)^2 \ \geq\ 0\,,
\end{eqnarray}
and hence
\begin{eqnarray}
    (D-2)H'(z)
    &\geq& \frac{(D-1)(D-2)}{4}\sum_{j=1}^{D-3} z\, f_{j}'(z)^2 \ \geq\ 0
    \quad\Longrightarrow\quad
    (D-2)H'(z)\geq 0\,.
\end{eqnarray}
\begin{figure}[t!]\centering
    \begin{subfigure}{0.45\textwidth}
        \includegraphics[width=\hsize,trim={0 0 -1.1cm 0},clip]{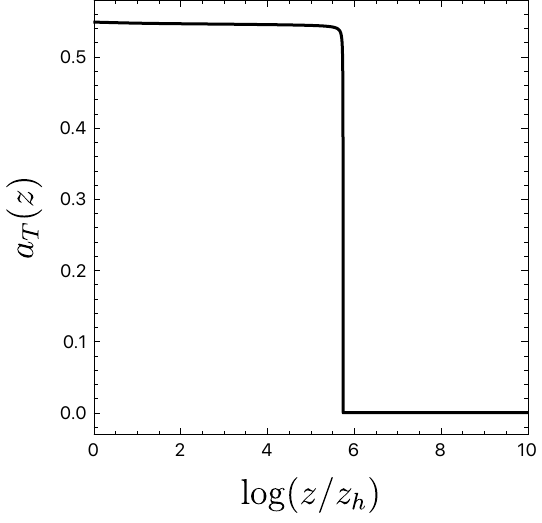}
        \captionsetup{justification=centering}
       \caption{}
        \label{fig:ae}
    \end{subfigure}
\hfil
    \begin{subfigure}{0.44\textwidth}
    \includegraphics[width=\hsize,trim={0 0 -1.1cm 0},clip]{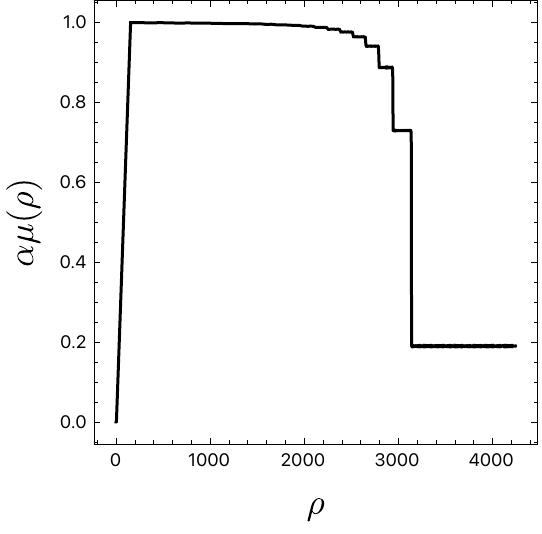}
    \captionsetup{justification=centering}
   \caption{}
    \label{fig:eons_ae}
\end{subfigure}
    \begin{subfigure}{0.44\textwidth}
        \includegraphics[width=\hsize,trim={0 0 -1.1cm 0},clip]{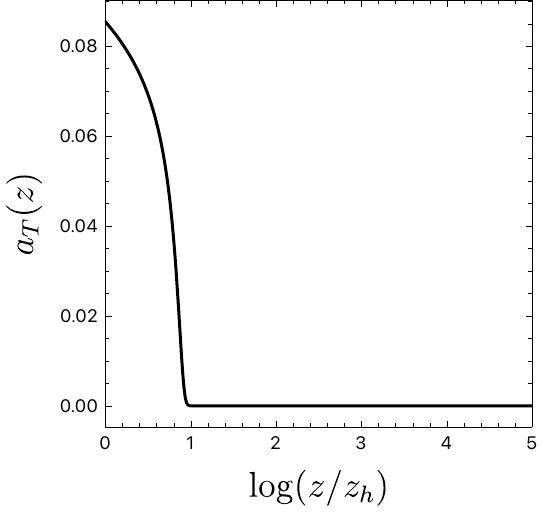}
        \captionsetup{justification=centering}
        \caption{}
        \label{fig:ac}
    \end{subfigure}
\hfil
    \begin{subfigure}{0.43\textwidth}
    \includegraphics[width=\hsize,trim={0 0 -1.1cm 0},clip]{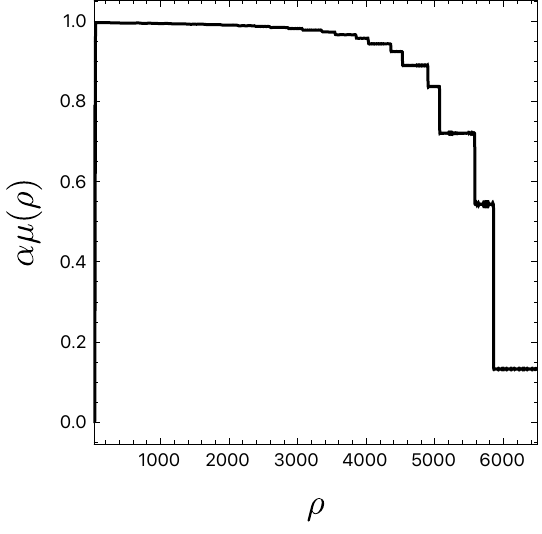}
    \captionsetup{justification=centering}
    \caption{}
    \label{fig:eons_ac}
\end{subfigure}
    \caption{Thermal $a$-function in units of $\pi^{(D-1)/2}/\ell_P^{D-2}\Gamma \left(\frac{D-1}{2}\right)$ (a and c) and its rate $\mu(\rho)=\frac{a'(\rho)}{a(\rho)}$ constructed to identify the Kasner epochs and eras (b and d) in $D=4$ and $D=5$. We nomalize $\mu(\rho)$ by $\alpha=-\frac{1}{(D-1)(D-2)}$ for convenience.}
    \label{fig:ae_ac}
\end{figure}
This immediately suggests the candidate thermal $a$-function to be
\begin{equation}
    a_T(z)=a_0\, e^{-(D-2)H(z)}\,,
\end{equation}
with $a_0=\frac{\pi^{\frac{D-1}{2}}}{\Gamma\big(\frac{D-1}{2}\big)\ell_P^{D-2}}$ a constant fixed by the UV value for the $a$-function. Since $H'(z)\ge 0$, we have $a_T'(z)\le 0$, so $a_T$ is monotone non-increasing along the entire radial evolution. In the trans-IR perspective, such monotonicity continues through the horizon and can extend all the way to Kasner-type singularities, where the effective number of degrees of freedom can vanish \cite{Caceres:2022smh}.

To resolve different interior epochs and eras i.e. \ sequences of Kasner regimes separated by bounces or chaotic transitions, it is useful to track the \emph{local decay rate} of $a_T$. Defining the logarithmic radial coordinate
\begin{equation}
    r=\log z\,,
\end{equation}
we introduce
\begin{equation}
    \mu(z)\equiv \frac{d\log a_T}{d\log z}
    = \frac{d\log a_T}{dr}
    = -(D-2)\,\frac{dH}{dr}
    = -(D-2)\, z\,H'(z)\,,
\end{equation}
which is non-positive by the NEC. In practice, $\mu$ provides a clean diagnostic: plateaus correspond to approximately constant Kasner exponents within an epoch, while sharp changes signal bounces between epochs. Such epoch-structured interiors arise in a variety of AdS black holes, including models with infinitely many Kasner epochs driven by scalar bounces \cite{Hartnoll:2022snh, Caceres:2023zft} and settings exhibiting never-ending Kasner alternation and Mixmaster-like chaos \cite{DeClerck:2023fax}.

Using the domain wall coordinate $\rho$ defined in (\ref{domainwallcoord}), we find that in the large-$\rho$ regime (deep interior), the function $H$ diverges linearly as
\begin{equation}
    H(\rho)\approx \frac{(D-1)\big(D-3+(D-1)p_0\big)}{2(D-2)}\,\rho\,,\qquad \rho\gg 1\,.
\end{equation}
Then the thermal $a$-function and its decay rate in a given Kasner epoch behave as
\begin{eqnarray}
    a_T(\rho) &\sim& \exp\!\left[-\frac{1}{2}(D-1)\big(D-3+(D-1)p_0\big)\rho\right]\,,\\
    \mu(\rho) &=& -\frac{1}{2}(D-1)\big(D-3+(D-1)p_0\big)\,,
\end{eqnarray}
so each epoch is characterized by a constant $\mu$ determined by the corresponding Kasner exponent $p_0$. Tracking $\mu$ across the interior therefore naturally segments the geometry into epochs/eras while preserving the global monotonicity of $a_T$ ensured by the NEC. Notably, when $p_{0}$ approaches the value $\tfrac{3-D}{D-1}$, the rate becomes nearly zero. From the CFT perspective, such regimes correspond to near-fixed points of the trans-IR RG flow, exhibiting approximately conformal (walking) behavior. Thus, even though the near-singularity regime is governed by intrinsically chaotic Kasner transitions, the probe \(a_T\) still defines a monotone along the interior flow. It remains well-defined throughout the evolution and provides a robust diagnostic that continuously tracks the geometry across successive epochs and bounces, without losing sensitivity to the underlying dynamics.\\
The $a$-function is stationary at the horizon as well as at the singularity,
\begin{eqnarray}
    &\frac{d a_T}{d\rho}\bigg{|}_{\rho=\rho_h}&=\frac{da_T}{d z}\frac{d z}{d\rho}\bigg{|}_{z=z_h}=a_T'(z_h)\sqrt{-F(z_h)n(\rho_h)}=0\\
    &\frac{d a_T}{d\rho}\bigg{|}_{\rho\to\infty}&=\mu(\rho)e^{-\frac{1}{2}(D-1)(D-3+(D-1)p_0)\rho}\bigg{|}_{\rho\to\infty}=0\\
\end{eqnarray}
In the last line, we used the inequality $p_0>\frac{3-D}{D-1}$. This ensures that both the horizon and the black hole singularity are the fixed points of the trans-IR RG flow. In particular, as $\rho\to\infty$ one finds
\begin{equation}
a_T(\rho)\longrightarrow 0\,,
\end{equation}
so the effective number of degrees of freedom measured by $a_T$ vanishes at the singularity, signalling a complete loss of degrees of freedom in this limit.


Using the bouncing rule for $p_0$ in \eqref{eq:kassol}, we can determine how the rate function $\mu$ transforms across a bounce. 
For a collision with the $m$-th wall ($m\neq 0$), the corresponding bouncing map acts on $\mu$ as
\begin{equation}
    \tilde{\mu}
    =\frac{-2\,(D-2)(D-1)\,p^{(0)}_{m}+(D-3)\,\mu}{D-3+2\,p^{(0)}_{m}}\, .
\end{equation}
For a collision with the $0$-th wall, the transformation instead becomes
\begin{equation}
    \tilde{\mu}
    =-\frac{(D-3)^{2}(D-1)\,\bigl(D^{2}-3D+2+\mu\bigr)}{(D-3)^{2}(D-1)-4\mu}\, .
\end{equation}
It is useful to interpret this dynamics as a sequence of bounces in a $(D\!-\!1)$-dimensional configuration space bounded by $(D\!-\!1)$ effective walls. Between successive collisions the trajectory is piecewise linear; in this sense, each straight line corresponds to a simple (monotonic) ``$a$--function'' evolution, while the reflections implement the discrete jumps \(\mu\mapsto\tilde{\mu}\) at the walls.

\begin{figure}
    \centering
    \includegraphics[width=0.45\linewidth]{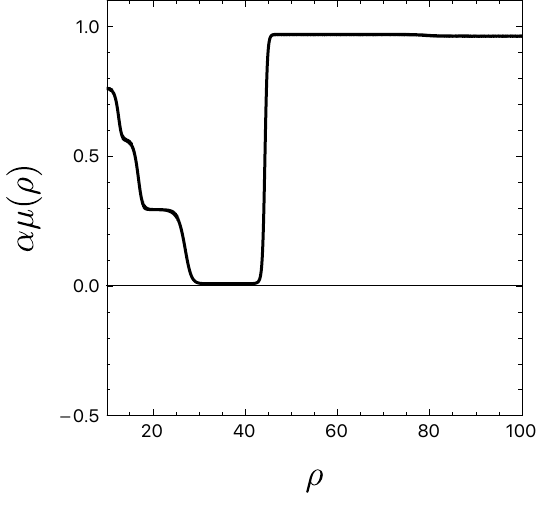}
    \includegraphics[width=0.425\linewidth]{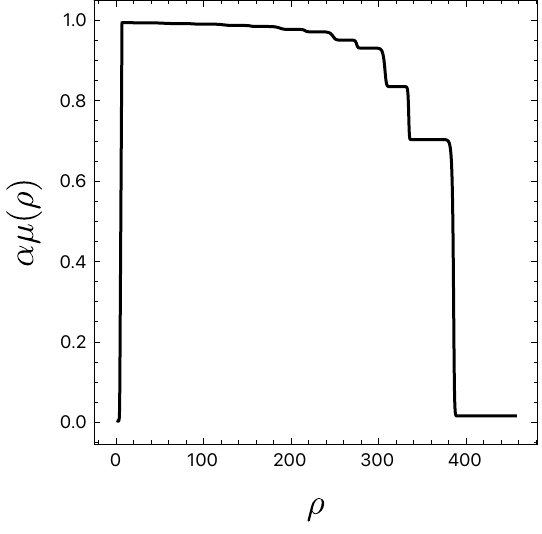}
    \caption{\textbf{Left:} Thermal $a$-function decay rate $\mu(\rho)$ along a representative Kasner-season-II trajectory in $D=5$. Approximately constant plateaus correspond to individual Kasner epochs, while the near-vanishing regime $\mu(\rho)\simeq 0$ signals the final (era-terminating) epoch. \textbf{Right:} The same quantity $\mu(\rho)$ for the symmetry-reduced dynamics confined to the triangular billiard, exhibiting an analogous epoch/era structure.}
    \label{fig:placeholder3}
\end{figure}

\section{Discussion}\label{sec:discussion}
In this work we have generalized the construction of \cite{DeClerck:2023fax} to asymptotically AdS black holes in arbitrary space-time dimension \(D\ge 4\). To the best of our knowledge, this provides the first explicit realization of bona fide BKL (Mixmaster) chaotic dynamics in higher-dimensional black hole interiors (for any $D \geq 4$). Our setup achieves this by coupling gravity to \((D-1)\) massive gauge fields in AdS, which generate the dominant electric walls that govern the near-singularity evolution and trigger the sequence of Kasner epochs and bounces characteristic of the BKL dynamics.

First, we have proven that such AdS black holes possess a single horizon in every $D \geq 4$, thus featuring a space-like singularity. Next, we have explicitly shown that the evolution towards the singularity in the deep black hole interior may be conceived as a sequence of Kasner regimes interconnected by very rapid transitions. In the appropriate parameter space, we have proven that these Kasner regimes correspond to free trajectories confined to the interior of a $(D-2)$-simplex, which suffer a \emph{bounce} as these hit one of the faces of the simplex. From this result, we were able to show the underlying chaotic nature of the subsequent Kasner dynamics for every $D \geq 4$. 

These aspects led us to the further scrutiny of the bouncing rules that connect the Kasner exponents of consecutive epochs in the deep black hole interior. A qualitative novelty that becomes manifest for \(D\ge 5\) is a richer internal organization of the Kasner evolution compared to the four-dimensional case. Beyond the standard epoch/era hierarchy, we find that successive epochs within an era can be connected by inequivalent transition maps, differing in the induced reordering of Kasner exponents. This motivated the introduction of \emph{Kasner seasons}, defined by the specific rules connecting a Kasner epoch with the successive one. We have explored in detail the five-dimensional case and fully characterized the subsequent electric-wall-driven BKL dynamics. We have observed that eras may be catalogued into various cases, depending on the pattern of Kasner seasons that the underlying Kasner epochs feature --- showing that three different Kasner seasons exist. We have proven that Kasner seasons are not an intrinsic property of electric walls, as these may also be identified in the chaotic dynamics associated to gravitational walls in five dimensions. While the bouncing rules corresponding to electric or gravitational walls are different in $D=5$, we were able to show that these are closely related, as applying twice the bouncing rules of electric walls yields those of gravitational walls, up to a permutation of exponents. We also initiated the study of BKL dynamics driven by electric walls in general dimensions $D>5$, providing the precise rules connecting the Kasner exponents and their reordering  between consecutive epochs, as well as showing the existence of $(D-2)$ Kasner seasons.


We also discussed a holographic diagnostic capable of identifying the interior epoch/era structure while remaining globally monotone: a candidate thermal \(a\)-function whose monotonicity follows from the null energy condition along a radially directed null congruence. Even though the interior undergoes chaotic Kasner transitions, the NEC guarantees that $a_T$ decreases monotonically. The corresponding decay rate $\mu$ is piecewise constant in each Kasner epoch and jumps at bounces, making it a useful diagnostic for identifying epochs and eras. Notably, as the relevant Kasner exponent approaches \(p_0\to (3-D)/(D-1)\), the rate becomes nearly zero, corresponding in the trans-IR interpretation to a near-fixed-point (``walking'') regime even though the overall flow continues toward vanishing \(a_T\) at the singularity.

These results open several natural directions. On the gravitational side, it would be natural to incorporate higher-curvature corrections and explore the interplay between epochs, seasons, eras, and eons in the interiors of asymptotically AdS charged black holes in higher dimensions \cite{toappear}. Also, in the same way that BKL dynamics in four dimensions is tantamount to the well-known Gauss map, it would be intriguing to study the explicit map encoding the chaotic behavior of Kasner dynamics driven by electric walls in general $D \geq 5$. Similarly, it would be interesting to systematize the season/era organization for $D>5$, both for electric- and gravitational-wall-driven dynamics. Specifically, one may ask whether the electric and gravitational bouncing rules remain connected for $D>5$. Such a connection should eventually break down, however, since the purely gravitational dynamics cease to be chaotic for $D>10$.

On the holographic side, it would be interesting to identify additional monotone probes that are not restricted to purely radial directions. Since BKL dynamics is driven by spacetime anisotropy, a diagnostic built solely from radially directed null congruences is naturally most sensitive to a particular Kasner exponent, namely $p_0$. In this work, we constructed such a globally monotone function using radial null vectors; a natural next step is to consider families of null vectors with components along anisotropic directions (see e.g.\ \cite{Giataganas:2017koz,Chu:2019uoh}). Such direction-dependent probes may remain monotone under appropriate energy conditions while being sensitive to different Kasner exponents, thereby providing a more complete holographic characterization of the anisotropic interior dynamics. Finally, it would be interesting to characterize BKL dynamics using quantum-information probes that can reach all the way to the singularity. In particular, various ``complexity=anything'' proposals~\cite{Jorstad:2023kmq,Arean:2024pzo} suggest observables that may remain sensitive to deep-interior structure even when standard geodesic or entanglement-based probes fail, and it would be worthwhile to explore whether they can resolve the sequence of epochs, seasons, and eras found here.


\section*{Acknowledgements}

We would like to thank Marine De Clerck, Sean Hartnoll, Robie Hennigar, David Mateos, Gerben Oling and Simon Ross for helpful and enlightening discussions. E.C. and \'{A}.J.M.  thank the Instituto de F\'isica Te\'orica UAM-CSIC for warm hospitality during the early stages of this work. The work of E.C. was supported in part by the National Science Foundation under Grant No. PHY–2210562 and by a  CNS-Spark grant (2025-2029). \'A.J.M.  was supported by a Juan de la Cierva contract (JDC2023-050770-I) from Spain’s Ministry of Science, Innovation and Universities. A.K.P. is supported by the European Union’s Horizon Europe research and innovation programme under the Marie Skłodowska-Curie grant agreement No. 101210745. J.F.P. was supported by the ‘Atracción de Talento’ program grant 2020-T1/TIC-20495, by the Spanish Research Agency through the grants CEX2020-001007-
S, PID2021-123017NB-I00 and PID2024-156043NB-I00,
funded by MCIN/AEI/10.13039/501100011033, and by
ERDF ‘A way of making Europe.’ 
\appendix

\section{Definition intervals of Kasner exponents}

Let us assume we have a set of $(D-1)$ real numbers $p_{D-2} \geq p_{D-3} \geq \dots \geq p_0$ satisfying the Kasner relations
\begin{equation}
    \sum_{i=0}^{D-2} p_i= \sum_{i=0}^{D-2} p_i^2=1\,.
\end{equation}
It is the purpose of this appendix to find the range of values that the different Kasner exponents $\{p_{D-2}, p_{D-3}, \dots, p_{0}\}$ may take. The strategy is simple. Assume $D-2\geq \lambda \geq 1$ exponents are equal to a real number $\alpha_\lambda$ and $D-1-\lambda$ exponents are equal to a different number $\beta_\lambda$. From the Kasner relations, it is clear that
\begin{equation}
    \alpha_\lambda^{\pm}=\frac{1}{D-1}\left (1\pm \frac{\sqrt{\lambda (D-1-\lambda)(D-2)}}{\lambda} \right )\,, \quad \beta_\lambda^{\pm}=\frac{1-\lambda\, \alpha^{\pm}_\lambda}{D-1-\lambda}\,, \quad D-2 \geq \lambda \geq 1\,,
\end{equation}
where we have incorporated the $\pm$ superindex to stress the fact that there are two independent solutions. A little thought reveals that the numbers $\alpha_\lambda^{\pm}$ actually correspond to the bounds of the various Kasner exponents  $p_{D-2} \geq p_{D-3} \geq \dots \geq  p_0$. As a matter of fact, it turns out that:
\begin{align}
\notag
& \alpha_1^{+}  \geq p_{D-2} \geq \alpha_{D-2}^{+}\,, \quad  \alpha_2^{+} \geq p_{D-3} \geq \alpha_{D-2}^{-}\,, \quad \alpha_3^{+} \geq p_{D-4} \geq \alpha_{D-3}^{-}\,,  \\  & \alpha_4^{+} \geq p_{D-4} \geq \alpha_{D-4}^{-}\,, \quad \dots \quad \alpha_{D-2}^{+} \geq p_{1} \geq \alpha_{2}^{-} \,, \quad \alpha_{D-2}^{-} \geq p_{0} \geq \alpha_{1}^{-}\,.
\label{eq:rangokas}
\end{align}
The bounds are optimal, since the exponents can actually take these values. We note that $1 \geq \alpha_\lambda^+ >0$, $0 \geq \alpha_\lambda^- \geq \frac{2}{D-1}$, $\alpha_1^+=1$ and $\alpha_{D-2}^-=0$. As a result, we may have a maximum of $D-3$ negative Kasner exponents and a maximum of $D-2$ positive Kasner exponents. 

Applying these results to $D=4$, we obtain the usual relations $1 \geq p_2 \geq 2/3$, $2/3 \geq p_1 \geq 0$ and $0 \geq p_3 \geq -1/3$. In $D=5$, one gets the bounds in \eqref{eq:rangoexp5}. Also, in $D=6$, one would get:
\begin{align}
\notag
    1 \geq p_4 \geq \frac{2}{5}\,,& \quad \frac{1+\sqrt{6}}{5} \geq p_3 \geq 0\,, \quad \frac{3+2\sqrt{6}}{15} \geq p_2 \geq \frac{3-2 \sqrt{6}}{15}\,, \\ & \frac{2}{5} \geq p_1 \geq \frac{1-\sqrt{6}}{5}\,, \quad 0 \geq p_0 \geq -\frac{3}{5}\,.
\end{align}

\bibliographystyle{JHEP-2}
\bibliography{Gravities.bib}

\end{document}